\documentclass[preprint]{aastex}
\usepackage{psfig}

\def\plotfiddle#1#2#3#4#5#6#7{\centering \leavevmode
    \vbox to#2{\rule{0pt}{#2}}
    \includegraphics{#1}}
\def\alwaysmath#1{\ifmmode{#1}\else{$#1$}\fi}

\lefthead{Majewski et al.}
\righthead{2MASS View of the Sagittarius Dwarf Galaxy }

\begin{document}
\title{A Two Micron All-Sky Survey View of the Sagittarius Dwarf Galaxy: 
II. Swope Telescope Spectroscopy of M Giant Stars in the Dynamically Cold Sagittarius 
Tidal Stream}

\author{Steven R. Majewski\altaffilmark{1},
William E. Kunkel\altaffilmark{2},
David R. Law\altaffilmark{1},
Richard J. Patterson\altaffilmark{1},
Allyson A. Polak\altaffilmark{1},
Helio J. Rocha-Pinto\altaffilmark{1},
Jeffrey D. Crane\altaffilmark{1},
Peter M. Frinchaboy\altaffilmark{1},
Cameron B. Hummels\altaffilmark{1},
Kathryn V. Johnston\altaffilmark{3},
Jaehyon Rhee\altaffilmark{1},
Michael F. Skrutskie\altaffilmark{1}, 
Martin Weinberg\altaffilmark{4}
}

\altaffiltext{1}{Dept. of Astronomy, University of Virginia,
Charlottesville, VA 22903-0818 (srm4n, drlaw, aap5u, hjr8q, jdc2k, pmf8b, cbh4r, 
rjp0i, jr2cz, mfs4n@virginia.edu)}

\altaffiltext{2}{Las Campanas Observatory, Casilla 601, La Serena, Chile
(kunkel@jeito.lco.cl)}

\altaffiltext{3}{Wesleyan University, Department of Astronomy, Middletown, CT
(kvj@urania.astro.wesleyan.edu)}

\altaffiltext{4}{Dept. of Astronomy, University of Massachusetts,
Amherst, MA 01003 (weinberg@astro.umass.edu)}

\begin{abstract}
  
We have obtained moderate resolution ($\sim 6$ km s$^{-1}$)
spectroscopy of several hundred M giant candidates selected from 
Two Micron All Sky Survey 
photometry.  Radial velocities are presented for stars mainly 
in the southern Galactic 
hemisphere, and the primary targets have Galactic positions consistent
with association to the tidal tail system of the Sagittarius (Sgr) dwarf 
galaxy.  M giant stars selected from the apparent
trailing debris arm of Sgr have velocities showing 
a clear trend with orbital longitude, as expected from models of the orbit and
destruction of Sgr.  A minimum 8 kpc width  
of the trailing stream about the Sgr orbital midplane 
is implied by verified radial velocity members.
The coldness of this stream ($\sigma_v \sim 10$ km s$^{-1}$) provides upper limits
on the combined contributions of stream heating by a lumpy Galactic halo
and the intrinsic dispersion of released stars, which is a function
of the Sgr core mass.
We find that the Sgr trailing arm is consistent with a Galactic halo that 
contains one dominant, LMC-like lump, however 
some lumpier halos are not ruled out. 
An upper limit to the total mass-to-light ratio of the Sgr core is 21 in
solar units. 

Evidence for other velocity structures is found among the 
more distant ($>13$ kpc) M giants.  A second structure that roughly mimics 
expectations for wrapped, leading Sgr arm debris crosses the trailing 
arm in the Southern Hemisphere; however, this may also be
an unrelated tidal feature.  Among the bright, nearby ($< 13$ kpc)
M giants toward the South Galactic Pole are a number with large
velocities that identify them as halo stars; these too may 
too trace halo substructure, perhaps 
part of the Sgr leading arm near the Sun. 

The positions and velocities of Southern Hemisphere
M giants are compared with those of Southern Hemisphere globular clusters 
potentially stripped from the Sgr system.  Support for
association of the globular clusters Pal 2 and Pal 12 with Sgr debris
is found based on positional and radial velocity matches.

Our discussion includes description of a masked-filtered cross-correlation
methodology that achieves better than 1/20 of a resolution element velocities in 
moderate resolution spectra.  The improved velocity resolution achieved 
allows tighter constraints to be placed on the coldness of the
Sgr stream than previously established.

\end{abstract}


\section{Introduction}

The extensive length of the tidal tails of the disrupting Sagittarius
(Sgr) dSph system has recently been demonstrated in all-sky views of this
system provided by the Two Micron All Sky Survey (2MASS) 
database (Majewski et al. 2003, ``Paper I" hereafter). 
The relatively metal-rich stellar populations in the Sgr
system means that M giant stars are prevalent in the Sgr debris stream.
These stars are easily identified to distances of more than 60 kpc using
2MASS $JHK_s$ photometry, and primary leading and trailing tidal arms from Sgr
are evident in all-sky M giant maps.
However, as described in Paper I, while Sgr debris
dominates the M giant population in the high halo, some ambiguities in the precise
length and placement of the Sgr arms remain due to contamination by other
M giant populations, particularly near the Galactic plane, and residual
photometric errors.  
Checks on the membership of M giants to the streams using radial velocities
can help delineate the extended morphology of the Sgr tidal tails. 

Stellar velocities are also useful tracers of mass.
As well as revealing clues to the 
structure and integrity of the Sgr core itself, the placement and motions of its
expansive tidal arms can provide important constraints on the mass of the Milky Way 
and the shape of its potential (e.g., Murai \& Fujimoto 1980, Lin \& Lynden-Bell 1982, 
Kuhn 1993, Johnston et al. 1999, Murali \& Dubinski 1999, Ibata et al. 2001, 
Law, Johnston \& Majewski 2004a,b).   
In addition, in principle, the velocity dispersion of tidal debris should 
provide a sensitive probe
of the lumpiness of the halo (Moore et al. 1999, Johnston, 
Spergel \& Haydn 2002, Ibata et al. 2002, Mayer et al. 2002).

The present study represents a first effort to accumulate velocity 
data on 2MASS-selected M giants in the apparent Sgr debris trails.
It includes a new radial velocity 
cross-correlation methodology that achieves better than 1/20 
of a resolution element discrimination; 
this analysis approach, in combination with the fact that M giants are 
intrinsically bright and accessible with modest telescope apertures,
places new, tighter constraints on the coldness 
of the Sgr stream using observations obtained with only a 1-m aperture telescope.  

We present results from a pilot radial velocity survey 
of 284 M giants from Paper I using the Swope 1-m telescope at Las Campanas
Observatory.  Most of the stars are in the Southern Galactic Hemisphere
and lie within 5 kpc of the best-fit Sgr orbital plane (Paper I).  
Forty percent of the stars were selected to be in the distinct, 
Sgr Southern Arc (trailing tail), and, as we show below, most of these
stars have velocities consistent with
that association.  The remaining stars are primarily very bright, nearby
M giants toward the South Galactic Pole, which were selected for study 
to probe the possibility that Sgr tidal debris may be quite
close to the Sun.  We show that this latter sample likely contains an admixture
of M giants from the Milky Way's Intermediate Population II/thick disk 
as well as stars from other halo substructure near the Sun --- perhaps
nearby Sgr stars.
Future papers will
detail observations obtained at other telescopes subsequent to 
the data collected and analyzed here.

\section{Radial Velocities of Sgr M Giant Candidates }

\subsection{Observing Program}

With one exception, 
M giant candidates selected for spectroscopic study have $0.95<J-K_s<1.15$.
and lie predominantly within 5 kpc\footnote{The 5 kpc separation is more
than 2.5 times the typical RMS separation of Sgr M giants from the nominal plane,
found to be 1.8 kpc in Paper I.  The M giant sample studied here should be 
representative of typical Sgr stream members across the width of the 
stream as viewed from the Sun.} 
 of the main Sgr plane defined in 
\citet{Majewski2003}, a plane consistent with a pole at $(l,b)=(273.8,-13.5)^{\circ}$.
A handful of observed stars are as distant as 6.3 kpc from that plane.  The spectra
for this project were taken as a backup to other observing programs (generally,
the Grid Giant Star Survey; Patterson et al. 2001).  Thus, the stars
for which spectra have been obtained were more or less randomly selected from
a much larger target list, and their distribution about the sky is a 
reflection of (1) available observing windows within the various observing 
runs for which time could be made for this program, and (2) a desire to
observe brighter M giants at any Sgr longitude\footnote{The 
Sgr coordinate systems used here are described at length in Paper I.
Briefly, the longitude angle $\Lambda_{\sun}$ is that angle, as viewed along the
Sgr plane from the Sun, from the Sgr center ($\Lambda_{\sun}=0^{\circ}$) and
increasing in the direction along the trailing Sgr tidal arm, which starts in the 
Southern Galactic Hemisphere.  Another parameter used frequently in this paper
is $Z_{Sgr,GC}$, which is the linear distance from the best-fit plane to the 
three-dimensional distribution of M giants in the Sgr tails.  }
when poor observing conditions 
drove observations to a bright star backup program.

All stars were observed with the Modular Spectrograph (ModSpec) of the Las Campanas 
Observatory mounted on the 1-m Swope telescope.  Candidate Sgr stars were observed
over the course of the observing runs listed in Table 1. 
The wavelength coverage extended from 4494\AA\  to 6720\AA\   with a dispersion of 
2.16 \AA\   per resolution element (about 113 km s$^{-1}$), but for the last 
two nights of the 2002 July 29-Aug 01 run the infrared 
Ca-triplet region was observed, with coverage between 7480\AA\  and 
9080\AA\  at a resolution of 1.51 \AA\ per resolution element 
(about 54 km s$^{-1}$).
The slit-width was set to the narrowest the instrument allowed, about 1.7 arcsec; 
yet a more or less ``smoothed out'' slit illumination 
results from image wander for most long exposures 
because of (1) seeing variations and (2) a slight polar axis misalignment
that produces slow drifts that are only periodically corrected out by the guider.
However, the data from 
2002 September 19-20 show a much larger scatter in repeat measures of radial
velocity standards and we believe that on those nights the slit had been 
accidentally jostled to a larger width for which the slit illumination was
not repeatable.  Fortunately, only a small number of
stars were observed on those nights.

During each run, multiple observations were obtained of multiple radial velocity 
standards to permit estimation of internal instrumental scatter and to calibrate
the zero-point of the velocity system (see \S 2.5).

\subsection{General Radial Velocity Cross-Correlation Methodology}

The methodology for deriving radial velocities is now being
utilized in a number of our research programs, but has not been previously 
described.
We set forth the details of these procedures here not only 
because of their wider application beyond the present survey, but 
because the improved velocity precision this
technique delivers is important to our analysis of the dynamics
of the Sgr debris stream. 

Several strategies are implemented to improve upon standard radial velocity
techniques and achieve a velocity precision better than 1/20 of a resolution element.
While the reduction to radial velocities employs essentially the classical 
cross-correlation methodology of Tonry \& Davis (1979), it is packaged 
in a pre-correlation editing process designed to address various 
challenges:

 One wants to eliminate from the input to the correlator all
spectral components that contribute no information.  Reliable kinematical
information is contained only in the slopes of unblended absorption
features that are known to be present in the range of spectral types
observed.  Therefore, the template input to the correlator is prepared from a high
signal-to-noise master from which the continuum is eliminated with a zero
phase-shift, high pass filter shaped to minimize sideband ``ringing''.
After filtering, the resulting spectrum is multiplied by a mask that is
zero everywhere except at a set of restframe wavelengths of low ionization
or low excitation transitions of elements observed in moderately metal
deficient stars;
these lines are taken from stellar atmospheric abundance studies of the anticipated
candidate spectral types, with
each feature represented by 
a Gaussian profile of unit amplitude and roughly 1.2 times the spectrometer
instrumental profile (about 2.5\AA\ ).  This mask is wavelength-shifted to
match the Doppler velocity of the master; a visual check for sideband symmetry
assures that no residual bias remains.  The low frequency portion of the
spectrum is discarded, leaving a working template with a mean of zero.
The target data stream is similarly high-pass filtered.  The one 
minor disadvantage of the prescription so far is 
that the high-pass filter leaves cosmetic noise (e.g., from 
CCD pixels or the rare cosmic ray) unmodified.

However, this approach brings an additional major advantage to our kinematic
endeavor.  Since the output of the correlator is affected only by the
strength of the selected spectral feature compared to that in the master, the
line list used may be tailored to respond well to a fairly broad range of
temperature, gravity, and abundance levels.  With spurious
or unknown features eliminated 
by the list
selection, we can abandon the requirement that a template match as
closely as possible the spectrum of the unknown candidate, and widen the
reach by editorially selecting those portions of an absorption trait for
which only two states are possible: either it is present, or it is not.
If present, it contributes by itself, and cleanly, to the correlation
peak;  if absent, what little damage is introduced from photon noise is
free of bias, and detracts minimally.  However, on abandoning the pursuit of
perfect matches in spectral typing, the cross-correlation function
is no longer a purely even function; thus odd terms no longer provide reliable
error estimation (see \S 2.4).

Despite the wide range of spectral types this approach
accommodates, it does fail when a spectral type is late enough that the
spectrum is dominated by strong molecular bands.  However, because the Sgr
candidates observed are generally of early M type, none  
had overly strong molecular features that made the velocity determination
fail, or even uncertain.
Another phenomenon inducing failure
is the presence of strong emission cores in selected absorption
features, as seen, for example, in Balmer line emission of late M
types and symbiotics, or Na D night sky emission in very faint spectra.  
Again, neither degradation affected the data set we describe.

\subsection{Details of the Master Cross-Correlation Template}

To build a masked, template master spectrum we start with a list of
lines for a stellar atmosphere that we believe to be approximately 
representative (in terms of log $g$, $T_{eff}$, [Fe/H]) of
our candidates. 
The choice of lines was guided by the stellar atmosphere studies 
of metal poor stars (similar to that expected of the candidates)
of Gratton \& Sneden (1988), McWilliam et al. (1995), and Ramirez \& Cohen (2003),
and included those spectral lines which, when compared to the same lines in the
Arcturus Atlas (Hinkle et al. 2000) were sufficiently
free of adjacent or blended features.  Additional clean lines not used by
those other studies but visible in the Arcturus spectrum were included.
We are interested in retaining parts of the spectrum that will
add power to the cross-correlation and rejecting those parts that
simply add noise.  By trial-and-error, and motivated by the typical 
exposure levels used in this investigation, a lower limit of
60m\AA\   equivalent width was adopted for retaining a line. 
The shape of the adopted mask at each line is 
a Gaussian profile of unit height and slightly wider than the instrument profile. 
This slight broadening of the profile ``softens" the effects on the 
cross-correlation of any slight miscentering of the {\it observed} lines
(e.g., due to jitter in the CCD pixel-to-log($\lambda$) transformation).

Finally, this mask is multiplied by a high signal-to-noise
spectrum of the star HD31871, a K0 III star, at the same resolution and same
instrument configuration as our target spectra.  Before 
this multiplication, the mask is shifted to the velocity of this star.

After experimentation with different templates for different observing runs, 
we found much better results by relying on the same template for the entire 
data set, because it minimizes sources of systematic error. 
Unsurprisingly, the sensitivity of a template designed in this way
varies with metallicity of the candidate, but it introduces no bias.  
Said another way, if a good cross-correlation is obtained, it is accurate.
While designed for [Fe/H] $\gtrsim -1.0$ K giant
stars, the final adopted template produced
reliable velocities for {\it all} Sgr stream target spectra having
typical exposure levels (which were about 300 electrons in the 
neighborhood of the Na D lines). 
Given our focus on {\it M giant} stars, which (for first ascent up the
giant branch) are produced in relatively metal rich populations,  
we did not encounter the severe metal depletions seen in our studies 
of halo K giants at large radial distances (e.g., Majewski et al. 1999) 
and that yield
only poor quality velocity estimates for
spectra exposed to more than 1000 electrons. 
Thus, the final adopted velocity cross-correlation template, which
was built relying on a line list limited to features wider than 60m\AA\ ,
proved ultimately to be a satisfactory match to the 2MASS M giant sample.

\subsection{Calibration and Reduction to Radial Velocities}

Calibration of the target stars relied on an observational procedure 
consisting of several independent steps.  
To provide a system calibration, 
prior to the start of each night of observing a
number of daytime spectra are obtained with the telescope at 
the zenith.  Exposures of a hollow cathode lamp combined with
He and Ne lamps provides a comparison spectrum yielding at least 30
comparison lines roughly evenly distributed over the wavelength range.  At
the beginning of some runs daytime spectra and companion comparison
lamp spectra were obtained at different
zenith distances and hour angles 
to confirm a satisfactory immunity from instrumental flexure. 

Nevertheless, comparison lamp spectra were taken for each star 
immediately following, and with the same airmass as, the target exposure.
Each candidate was observed in a single exposure to at least 300
electrons, with multiple exposures obtained only when an immediate quick-look
inspection afterward prompted a second exposure because of concern for 
the signal-to-noise in the first exposure. 
However, in many cases, both of the individual exposures proved more than adequate
for a good velocity measurement, and these
repeat exposures can be used as an integrity check on instrumental precision
(see \S2.5 below).  
It is worth noting that for observations in the
Ca-triplet domain longer than 300 seconds a superior wavelength calibration was
obtained from employing approximately
62 lines of the simultaneous night sky spectrum; 
a line list for this purpose, edited to avoid blends for the instrumental resolution,
was obtained from Osterbrock et al. (1996).

 Spectral extraction included the manual suppression of local
cosmetic defects; this proved more reliable as well as much more (observing) time
efficient than multiple exposures combined by median averaging.  From 
experience over years of using ModSpec
at the 1-m telescope we have found that a star drift of a fraction of a
pixel between exposures along the slit invalidates the presumption that
median averaging yields good results.
The origin of the slit
drift is a combination of change in atmospheric dispersion and in flexure
of the external guider probe.

The cross-correlation methodology described in the previous subsections was applied
to both target stars and numerous radial velocity standard stars observed
during each observing run.

\subsection{Error Analysis}

       The estimation of velocity uncertainties relies on an empirical 
calibration of scatter in a set of comparable velocity measurements obtained 
over years of using this methodology.  Basically, the strength of the correlation
peak, as well as its shape and degree of asymmetry, are compared in a
sequence of differently exposed spectra of known velocity, or,
equivalently, by degrading high quality spectra by adding pseudo-counting
noise.  The degradation is reflected in the visual appearance of the
correlation peak in relation to its sidebands.

       Since it is not practical to degrade every spectral exposure to
ascertain the robustness of an estimate, we summarize the results of
experiments with which an acceptance threshhold for good measures was 
devised, based on experience with many thousands of spectra obtained in 
more than forty observing runs.  In all cases we inspect the
cross-correlation spectrum over a range of values corresponding to
reasonable velocity values for Galactic stars (roughly $\pm600$ km s$^{-1}$).
The acceptance threshhold for a good spectrum relies on the appearance of the
normalized cross-correlation peak, which must fulfill two criteria to be
satisfied in a visual inspection: (1) The nearest sidebands in the 
cross-correlation spectrum must be no greater than 60\% of the correlation peak, and
(2) the FWHM of the central correlation peak must be either fully symmetric above
zero, or, if an asymmetry is seen, this must not appear at a point in the
central peak that is more than 50\% 
the maximum value of that peak.  It must be noted that the strength
of sidebands in the correlation plot varies with the number of features in a line
list from which the velocity template is built.  When few features build a template,
sideband strength increases.  Consequently the criterion described here is
specific to the template described in \S2.3.

Objects just marginally satisfying the above criteria generally reflect a velocity 
uncertainty less than double that obtained when nominally perfect correlation 
peaks are obtained.  But when the strongest correlation sideband is $\le$50\% 
the height of the central peak the velocity precision is no worse than 
the optimal precision in 95\% of the cases.
When asymmetries are seen only in the sidebands,
the candidate spectral type is different from the template.  However, comparison
with velocity standards covering a range of spectral types has failed to impute
a velocity bias in these cases.

        To summarize our evaluation of the appearance of the correlation
peak and sidebands for each spectrum, we have devised a quality scale from $Q=0$ (worst)
to $Q=7$ (best) that we provide with the data in Tables 2 and 3.\footnote{ The 
most reliable spectra, with perfectly
symmetric sidebands, are classified as $Q=7$, while a cross-correlation plot with
random noise high-pass filtered like a real spectrum is classified as $Q=0$.
Other $Q$-values assigned carry the following significance: A $Q=6$ shows some
deviation in the sideband distribution, with one side or the other showing 
stronger sidebands.  
When sideband symmetry is retained, but the strength of 
correlation peak declines to where the strongest sidebands reach half the 
peak amplitude, $Q$ is set to 5; in practice there is no sensible loss in the
reliability of the velocity precision.  $Q = 4$ is the lowest quality estimate
that is reliable, and scatter from degradation simulations appears to be about
double that seen in $Q=7$ data.  
$Q=3$ is assigned when the strongest sideband 
rises to about 70\% of the correlation peak;  summing two $Q=3$
spectra tends to give a $Q=4$ spectrum.  A value of $Q = 2$ is assigned when there
is still no ambiguity as to which peak is the correlation peak, and the 
summing of two $Q=2$ spectra yields improvement (just barely) in the appearance 
of the correlation plot.  In practice the velocity estimates are discarded,
however.  $Q=1$ is assigned to such data where there is some indication that a
heftier exposure might produce signal improvement (as would not be the case, for example,
when correlating a quasar spectrum with a K7 III template).  }
We present here only stars with $Q \ge 4$, but only four stars have qualities 
at this limit, and only one of these is in the trailing arm, which is the focus 
of attention in this paper.  An additional parameter describing the quality of 
the radial velocities is the peak of the cross-correlation spectrum ($CCP$).  
Both $CCP$ and $Q$ are internal
descriptors that allow quantitative relative comparison within our instrumental
system peculiar to our instrumental setup.  Significance does not attach
to either number outside of that context.  The $Q$ index is most useful for 
descriptions of spectra with ``borderline" $S/N$, because the size of the CCP
is a function of the number of lines allowed by the mask as well as the strength
of lines in the target spectra: For example, the ten stars presented in
Table 3 with $4<Q<6$    
have $0.21 < CCP < 0.43$, while the $Q=7$ stars span $0.29 < CCP < 1.25$.

Among stars that have met our acceptance criteria are the sets of dozens
of radial velocity standards with spectral types spanning those of targets 
observed during each observing run.  The scatter in derived velocities among
these standards are one indicator of the precision of the velocity system. 
Table 1 demonstrates the derived velocity scatters of velocity standards
for each observing run on which Sgr M giant candidates were observed.  
However, we believe these scatters to be {\it upper limits} to the true random 
errors:  During this and previous observing  
projects it was learned that the brighter standards in the range ($ 2.0 < V
< 10.0$) occasionally yielded large deviations; investigation of this problem
revealed it to arise from stellar profiles that illuminated the slit 
incompletely during very short exposures with respect to the seeing timescale 
(as short as one second in some cases).  Indirect
confirmation of this interpretation stems from the observation that nights
of best seeing yielded increased 
calibration scatter.  Nevertheless,
while we believe the true random errors may approach 
4 km s$^{-1}$ in some cases, we adopt a typical mean random error of 6 km s$^{-1}$ 
to be the most representative value over the entire spread of spectral types
and over all observing runs.
This does not apply to the twelve Sgr candidates observed 2002 September 19-20 
with the slit problems discussed in \S2.1, and for which larger errors are expected. 

It may be argued that the radial velocity standards, which typically are
bright and yield excellent signal-to-noise spectra, will produce estimates of the
velocity scatter that are unrepresentative of expectations
for the fainter M giants.  However, we find similar dispersions about
the mean value for those M giants with multiple observations.  These
multiple observations and their dispersions are summarized in Table 2.  
In most cases the repeat measures are taken on the same night (indeed,
in succession --- see the ccd frame numbers given in Table 2), but there
are exceptions when the same star is occasionally observed on different nights.
In these cases, night-to-night consistency is found, which suggests that
any inherent, systematic errors are small.  This is so even 
in two examples when different spectrograph set-ups were used (marked by
footnote ``d" in Table 2).

The more numerous radial velocity standards provide a better gauge of 
{\it systemic} errors in the velocities produced from our manufactured velocity template.  
After compensating for deviations in the very brightest standards taken with
the shortest exposures, we cannot reject the hypothesis that a
unique velocity offset validly describes an entire observing run, even for
those runs with the largest set of standards observations and spanning spectral types
as early as G5II and as late as M1.5III.  Indeed, from run to run we rather consistently 
determined that a velocity offset of 12 km s$^{-1}$ corrected our template 
velocities to the IAU standard.  Consequently a single velocity offset 
(typically around 12 km s$^{-1}$), derived from observations of the radial velocity standards, 
was applied to cross-correlation velocities derived for all stars in each 
separate observing run.

\section{Velocity Distribution with Orbital Longitude}

All stars observed yielded spectra that contain some
molecular line blanketing, as expected for late-type stars, and have a general appearance
of early M types.  This is in keeping with the general color range of most of the targets 
(typically $0.95 < J-K_s < 1.10$), which is centered on those of early M giants (see,
e.g., Fig. 21b in Paper I).  
Spectroscopic discrimination of gravities (dwarf/giant separation) was not attempted,
because the stars have been preselected to be giants based on the 2MASS 
($J-H, J-K_s)_o$ two-color diagram.
At the magnitudes and colors probed here ($6 < K_s < 12.5$) any late-type dwarf 
stars ($M_K > 7$) creeping into the sample
would be within $\sim100$ pc of the Sun, and typically be closer than 50 pc.  
Such stars would be expected to have orbits like the Sun and
therefore have low heliocentric radial velocities, but potentially
large proper motions.  However, a match of all Table 3 M giants
to the UCAC2 catalogue (Zacharias et al. 2001) shows none of the
stars to have proper motions larger than 30 mas year$^{-1}$, most 
to have motions more like 10 mas year$^{-1}$, and all
to have reduced proper motions solidly indicative of giant stars. 
%
%
 The overall reliability of the
color-selection for M {\it giants} is further demonstrated by the relatively
clean Sgr sample exhibited below (see Fig. 2a) even for the faintest, most 
``vulnerable" magnitude range probed.

Table 3 gives the radial velocity measurements for the first 284 M giant candidates observed
in this program with the Swope telescope.
For stars observed multiple times, the Table 3 entries represent averages as shown in 
Table 2 (the rules adopted for weighted averaging are given in the Table 2 notes).
The columns of the table are the star name (derived from the Equinox 2000.0
coordinates)\footnote{These star names have been generated with our own software
and may differ slightly from the official names for the stars in the 2MASS database
due to differences in rounding the coordinates.},
dereddened $K_s$ magnitude and $J-K_s$ color (the dereddening procedure is
described in Paper I), Galactic coordinates in degrees
($l,b$), and, from Paper I, the photometric parallax distance ($d$) in kiloparsecs, 
the Sgr longitude $\Lambda_{\sun}$, and distance from the best fit Sgr plane ($Z_{Sgr,GC}$) in
kiloparsecs.
This is followed by the measured heliocentric
($v_{hel}$) and Galactocentric Standard of Rest radial velocity ($v_{GSR}$)\footnote{A 
Local Standard of Rest rotation velocity of 220 km s$^{-1}$ and a solar
peculiar velocity of ($u, v, w$)=$(-9,12,7)$ km s$^{-1}$ are adopted.} in km s$^{-1}$, 
the peak of the cross-correlation
curve ($CCP$), cross-correlation quality index ($Q$), and the date of the observation.
Figure 1 shows the planar and longitudinal distribution of stars for which radial velocities
have been obtained.
Three primary areas have received attention thus far: (1) Southern Hemisphere
stars with positions consistent with the trailing tidal debris (e.g., the more distant stars on
the right side of Fig. 1e and towards the bottom of Figs. 1a and 1b), (2) nearby
M giants towards the South Galactic Pole (e.g., the tight clustering of nearby stars 
in the lower of Figure 1e), and (3) a smattering of 
M giants in the Northern
Galactic Hemisphere (e.g., $Z_{GC}>0$ stars in Figs. 1a and 1b and on the left side
of Fig. 1e), predominantly nearby stars towards the North Galactic Pole.
A few stars in the more distant part of the Northern Loop of the leading debris
arm have also been observed.

\subsection{Distant Southern Hemisphere M Giants}

\subsubsection{General Character of the Sgr Trailing Stream Velocities}

Figure 2a shows the radial velocity distribution of the Southern Hemisphere stars more
than 13 kpc from the Sun as a function of orbital longitude.
The primary feature demonstrated is the smoothly
varying radial velocity along the Southern Arc extending from the Sgr core
(the latter represented by the positions of the Sgr globular clusters in Fig. 2a ---
see also \S5), which shows the approach of this
trailing debris in the direction of the outer Galaxy ($\Lambda_{\sun} \gtrsim 75^{\circ}$)
and its net recession from us after it passes through the South Galactic Cap
(near $\Lambda_{\sun} \sim  75^{\circ}$).  The recessional velocity of this tidal 
debris eventually approaches that of the Sgr core (171 km s$^{-1}$
as derived by Ibata et al. 1997)
and core globular cluster system (M54, Terzan 7, Terzan 8, Arp 2) at $v_{GSR} \sim
+175$ km s$^{-1}$.
The largest approaching $v_{GSR}$ velocities towards the direction of the Galactic anticenter
(at roughly $\Lambda_{\sun} \sim 165^{\circ}$) approach $-200$ km s$^{-1}$.  

As expected for energy-sorted debris from a disrupting satellite, the Sgr stream is
extremely velocity coherent.  
In the region of the South Galactic Cap and sweeping towards the Sgr core
our line of sight nearly perpendicularly traverses the Sgr trailing arm 
(see Figs. 10 and 11 of Paper I, wherein it may be seen that the Sgr 
stream arcs almost equidistantly around the solar position for 
$\Lambda_{\sun} < 100^{\circ}$), and in this region we can, in principle, 
most directly measure the energy spread of the debris.  A third order 
polynomial fit (utilizing a 2.85$\sigma$ rejection\footnote{We desire to
``fairly" remove stars that are clearly unrelated to the main Fig. 2a trend. 
With an iterative $\sigma$-rejection algorithm, it turns out that {\it no} stars 
are removed until a 2.9$\sigma$ limit is adopted and we
adopt a 2.85$\sigma$ rejection as most representative of what we believe
the typical person would reject as an outlier 
(see filled boxes in Fig. 2a).}) to the velocities of 
debris stars from $25^{\circ} < \Lambda_{\sun} < 90^{\circ}$ yields a 
dispersion of 11.7 km s$^{-1}$ about the following mean trend for
$v_{GSR}$ in km s$^{-1}$ as a function of $\Lambda_{\sun}$ in degrees (for 60 stars, 
with 15 iteratively rejected and ignoring one observation from 
2002 September 20 --- see Fig. 2a for the stars used in the fits described in this
section):

$$ v_{GSR}=270.0 - 5.221 \Lambda_{\sun} + 0.01520 \Lambda_{\sun}^2$$

\noindent The adopted lower longitude 
limit is just inside the nominal King profile limiting radius {\it along 
the major axis} of Sgr as derived in Paper I, and this limiting radius 
should be an upper limit to the tidal radius of the satellite.
Some fraction of the observed dispersion (Fig. 3) is contributed by 
observational error.  The 6 km s$^{-1}$ estimated typical random error (\S2.5) 
removed in quadrature leaves an intrinsic stream dispersion of 10.0 km s$^{-1}$.  
The uncertainty in this derived intrinsic dispersion is 1.4/1.7/2.2 km s$^{-1}$ if we 
assume an uncertainty in the per star random error of 1/2/3 km s$^{-1}$;
we adopt the middle values as appropriate (and, except where noted, we use
2 km s$^{-1}$ as the uncertainty in the velocity uncertainties to 
calculate error in dispersions) and thus find that the
intrinsic dispersion is $10.0 \pm 1.7$ km s$^{-1}$.

Were one instead to adopt optimistic (say $4\pm1$ km s$^{-1}$) to pessimistic 
($9\pm1$ km s$^{-1}$) estimates of the mean random error, intrinsic stream 
dispersions of $11.0\pm1.2$ and $7.5\pm2.1$ km s$^{-1}$, respectively, are obtained.  
If we adopt for the uncertainty of each star the date-specific scatter 
values given in the second column of Table 1 
we obtain a mean uncertainty of 5.3 km s$^{-1}$, which yields an intrinsic
stream dispersion of $10.4\pm1.3$ km s$^{-1}$.  For comparison, the measured 
velocity dispersion of the Sgr core, based on the ``f7" field of Ibata et al.
(1997), is $11.4\pm0.7$ km s$^{-1}$.
We conclude that  
for the range $\Lambda_{\sun}=25-90^{\circ}$, which corresponds primarily to debris
lost within about the last half-orbit of Sgr ($<0.4$ Gyr; see Law et al. 2004a,b),
the trailing tidal arm stream is only slightly colder 
than the core of the parent dSph.\footnote{One other astrophysical effect
that acts to {\it increase} the observed velocity dispersion has been ignored here:
Stars near the tip of the giant branch show an intrinsic atmospheric velocity jitter.
The amplitude of this velocity jitter appears to increase with decreasing 
surface temperature or increasing luminosity (e.g., Pryor et al. 1988, C\^ot\'e et al. 1996,
Frink et al. 2001) and should be most intense for
M giants, perhaps exceeding 1000 m s$^{-1}$.  
Thus, radial velocity dispersions of populations estimated from measurements of 
M giants should slightly {\it overestimate} the true dispersion of the 
population (Olin Eggen 1970, private communication to WEK; also, Gunn \& Griffin 1979).}

However, Figure 2a shows an apparent increase in the velocity spread with 
longitude, perhaps even a ``jump" in dispersion at 
$\Lambda_{\sun} \sim 90^{\circ}$.  This ``jump" is not a result of 
a change in the mean error of the measures at this point, because, if we 
adopt the 
scatter values in Table 1 to assign errors to each star, the mean uncertainty
in this group rises only to 6.2 km s$^{-1}$.
For $90^{\circ} < \Lambda_{\sun} < 145^{\circ}$ the observed 
dispersion increases to $14.2$ km s$^{-1}$ (for 44 stars after rejecting the
four stars near $\Lambda_{\sun} = 135^{\circ}$ and 0 km s$^{-1}$ and
ignoring four stars from 2002 September 20),  
so the true dispersion does indeed appear to be larger (see Fig. 3) for 
this longitude range: $12.8\pm1.9$ km s$^{-1}$.
Including the data from the entire range
$90^{\circ} < \Lambda_{\sun} < 150^{\circ}$, the observed dispersion is 
13.6 km s$^{-1}$ (with 2.85$\sigma$ rejection on a sample of 108 stars, 
with 19 iteratively rejected, and ignoring
five stars from 2002 September 20
observations)\footnote{This dispersion is that about the best fitting 
polynomial $ v_{GSR}=283.1 - 5.594 \Lambda_{\sun} + 0.01725 \Lambda_{\sun}^2$
km s$^{-1}$, where $\Lambda_{\sun}$ is in degrees.},
which translates to an intrinsic dispersion of $12.3\pm1.3$ km s$^{-1}$.   

The increase in velocity dispersion may reflect a number of aspects 
intrinsic to the evolution of the Sgr arms in a smooth halo potential:  
First, a {\it gradual} increase
in velocity dispersion might be expected simply from the fact that at larger
$\Lambda_{\sun}$ our line of sight to the trailing tail becomes more oblique,
so the observed ``depth'' and velocity dispersion has an increasing contribution 
from the dimension {\it along} the tail.  Second, debris released from
different perigalacticon passages of Sgr can overlap, putting stars of
different orbital energies at the same orbital phase ($\Lambda_{\sun}$).  
The effect is demonstrated in the models we present in Law et al. (2004a,b), 
where, indeed, some overlap of debris from one and two orbits ago does begin
in the larger $\Lambda_{\sun}$ range discussed above.  Third, 
because Sgr is ``crumbling", its mass, and therefore presumably its internal velocity
dispersion, must have been larger in the past, so that the dispersion 
should increase in older debris more distant from the Sgr core.  Finally,
because stars spend more time in the outer parts of their
orbits, one might expect a gradual ``piling up" of stars of all 
energies at these parts of the debris orbits. 
Each of these effects might be expected to contribute to an inflation
of the dispersion at larger $\Lambda_{\sun}$.\footnote{With sufficient
velocity resolution, a {\it bimodal} velocity distribution might be
observable as a result of the second effect discussed.}  A fifth effect, 
the heating of the debris by dark matter lumps, is addressed in \S4.

In conclusion, over the full length of the trailing arm as traced by our 
M giant velocity data, the Sgr debris is very energy coherent, of order,
or slightly smaller than, the dispersion of the actual satellite core, but with 
a velocity dispersion that increases with longitude.

\subsubsection{Comparison to Previous Work}

It is worth comparing these results to the one other published measure of
Sgr radial velocities in the trailing arm tail by Yanny et al. (2003).  Through 
the Sloan Digital Sky Survey, they have measured the velocities of 
apparent Sgr main sequence turn-off (MSTO) stars in a field near  
$\Lambda_{\sun} \sim 110^{\circ}$.  The mean velocity of their Sgr stars, 
$-89$ km s$^{-1}$, is shown in Figure 2a (after the same conversion from 
$v_{hel}$ to $v_{GSR}$ as we have made for the M giants), and is seen to 
be offset by about $+40$ km s$^{-1}$ from the mean of our own velocities at 
this same point of the Sgr stream.  Either the Yanny et al. sample of MSTO 
Sgr stars is displaced to a different energy from the M giant Sgr sample  --- a 
rather unlikely situation --- or one of the two surveys contains a 
systematic error in their velocity system.  

We have reason to believe the veracity of the M giant velocity results
presented here.  First, our M giant velocity trend extrapolates nicely to those
radial velocities previously observed in the Sgr core for both stars (Ibata 
et al. 1997) and the four primary globular clusters (Fig. 2a).  Second, 
our reduction methodology has been in use for years on a variety of programs 
led by one of us (W.E.K.) that have yielded numerous external checks on the code and procedures.  
In addition, apart from radial velocity standard stars discussed earlier, 
over the course of our GGSS and Sgr observations
we have observed over a hundred metallicity/gravity standards (some two dozen on
the nights of Sgr observations); of these, more than 2/3 have published velocities
against which no systematic velocity error was found.  And finally, despite the facts that 
(1) our velocities derive from one-at-a-time measures of individual stars
over the course of the six observing runs (Table 1), (2) two different spectral regions
were used over the course of those runs, and (3) each run was reduced separately 
and checked against generally different mixes of multiple radial velocity standard
stars, the observed velocity {\it coherence} of the M giant
Sgr trailing stream (Fig. 3) is within a factor of two of that expected
from the per star {\it random errors} of our velocity measures.  
Were our {\it systematic} errors of order the size of the observed velocity 
offset between our data and that of Yanny et al. (2003), whether from use of 
bad velocity standard stars (but we emphasize that the specific ones used generally varied 
from run to run) or from any more basic problems related to the 
spectrograph set-up (e.g., varying slit width, placement of the spectrum on the CCD, 
or wavelength range observed), one might expect variation in such systematics from 
observing run to observing run, and these variations would likely grossly inflate
the dispersion of the Sgr stream beyond what is observed.  

A separate comparison to the Yanny et al. (2003) survey is possible via
velocity trends found for the ``Monoceros" anticenter structure that is the 
focus of their work.  In a separate analysis of M giants 
in the same region of the ``Monoceros" structure, 
with M giant observations from a different telescope and using different 
reduction software, Crane et al. (2003)
find a $20$ km s$^{-1}$ offset (in the same sense as for the Sgr comparison)
compared to the SDSS MSTO stars.  

Yanny et al. also find a 21 km s$^{-1}$ observed Sgr stream velocity
dispersion, which they correct to a true dispersion of 20 km s$^{-1}$ after 
accounting for a 7 km s$^{-1}$ estimated random error for their stars.  
This corrected dispersion is significantly larger than our {\it observed}
14.1 km s$^{-1}$ velocity dispersion for Table 3 M giants at a
similar point in the Southern stream 
(defined for present purposes by 17 M giants in the range $100^{\circ} 
< \Lambda_{\sun} < 120^{\circ}$).  
Both surveys report similar radial velocity uncertainties ($\sim 7$ km s$^{-1}$
by Yanny et al., $\sim 6$ km s$^{-1}$ for the M giants), but the 
M giant spectra were taken with twice the resoluton and at higher signal-to-noise. 
Better agreement in the derived true Sgr stream velocity dispersions 
would be found by increasing the SDSS velocity errors by a factor of two or more.  
However, even increasing the {\it random} errors of either survey by a factor of 
two still does not explain the mean velocity {\it offset} between them.

\subsubsection{The Width of the Sgr Trailing Stream}

An issue pondered 
in Paper I is that of the breadth of the Sgr tidal streamers on the plane of the sky.  
Based on the best fit plane to distant M giants, 
Paper I found a 1-$\sigma$ spread of likely Sgr stars
about that plane of $\lesssim 2$ kpc, but with stars seeming to spread from the
best fit plane to several times more than that.
The coherent grouping of stars that, based on their radial velocities, are obviously 
part of the trailing debris tail provides a useful check on this earlier assessment.
Though this debris stream is far from completely sampled, a number of stars
as far as 3 kpc on one side of the plane ($Z_{Sgr,GC} = -3$ kpc in the
Galactic Center coordinate system introduced in Paper I) and to 5 kpc
on the other side ($Z_{Sgr,GC} = + 5$ kpc) are radial velocity members
of this stream.  This stream dimension is only weakly (and linearly)
dependent on uncertainties in the distance to the stream (which are about 17\%).  Thus, the
width of the trailing stream on the plane of the sky would seem to be at least
8 kpc, or 10 kpc if symmetry about the $Z_{Sgr,GC}$ midplane is invoked.
This width  is comparable to the minor axis radius of the Sgr main body
from Paper I --- about 5 kpc.

The $8-10$ kpc breadth, if similar to the {\it depth} of the
stream in line of sight, suggests that the Sgr debris streamers cut a relatively
large swath through the Galaxy.  In turn, this supports the view that Sgr debris 
from the {\it leading arm} may be in the solar neighborhood, despite 
uncertainty on exactly where the center of this locus traverses the disk, {\it as long as}
this tidal arm extends far enough to reach the Galactic plane on this side
of the Galactic Center.

\subsection{Other Tidal Features Among the Distant Giant Sample?}

Outliers from the Sgr locus in Figure 2a present tantalizing clues to
other possible tidal features in the halo.  
Analysis in Paper I (see \S 6.5 of that paper) 
of possible contamination of the trailing M giant tidal arm 
suggested that about 10\% of the M giant candidates selected at each 
longitude could be stars unrelated to Sgr tidal debris.  Yet at many longitudes
the radial velocity results here yield a success rate better than that assessment.
For example, over the range $70^{\circ} < \Lambda_{\sun} < 125^{\circ},$ only
two or three of the five dozen or so stars have velocities far enough away from the Sgr
locus that it is conceivable they are unrelated --- but even some of these may just
be displaced by observational error (all are $Q=7$ spectra but several of the 
less extreme outliers are from the 2002 September 19-20 observing run that had the 
slit width problems). 
On the other hand, the vast majority of the handful of the 127 distant M giants shown in 
Figure 2a that are most 
inconsistent with Sgr debris are concentrated to particular  $\Lambda_{\sun}$ ranges, 
so that only certain points of the Sgr trailing tail show significantly larger numbers of 
``contaminating" stars off the main velocity locus.  All of these more extreme outlier
stars have $Q=7$ spectra from reliable observing runs, so their velocities 
do not likely reflect large observational
errors.  A random distribution of contaminants might not be expected to vary
wildly in density with position along the Sgr tail ($\Lambda_{\sun}$).   
Indeed, the contaminating stars
seem to have their own velocity coherence, and these velocity trends tantalizingly suggest
the existence of other phase space clumpings among the M giants in the halo.

Prominent among the outliers are four\footnote{Stars 0322152+164323,
0358367+071547, 0403354+055238, and 0422084+083418.}
at $\Lambda_{\sun} \sim 135^{\circ}$ 
having $v_{GSR}$ from $-35$ to $+5$ km s$^{-1}$.   These stars are at $b \sim -30^{\circ}$ 
and $l \sim 180^{\circ}$.  Interestingly, this is the general direction of the ``ring" feature 
(Ibata et al. 2003, Yanny et al. 2003, Rocha-Pinto et al. 2003) identified with the
potential ``new galaxy" or tidal feature identified as S223+20-19.4 by Newberg et al. (2002).
This ``Monoceros" feature (as referred to by Yanny et al. 2003) is also visible in
the 2MASS data as discussed in \S 8.1 of Paper I (see Fig. 19 of that paper,
and also Rocha-Pinto et al. 2003)
and the four stars above have a mean velocity typical for that feature (see Crane et al.
2003).  Unfortunately, the photometric parallax distances for these four stars ($17-30$ kpc)
are too large for association with the Newberg et al. (2002) anticenter
feature (see Fig. 1 of Crane et al. or Fig. 4 of Rocha-Pinto et al. 2003),
and would require the adoption of a more metal-rich color-magnitude relation 
to bring them closer; yet Crane et al. report evidence that the 
M giants in the anticenter structure are, if anything, slightly more metal {\it poor}
than those in the Sgr stream.  Even more enrichment would be required to put these 
four stars 
at distances expected for the Intermediate Population II (IPII) thick disk stars, but this 
would these four stars substantially more metal rich 
than expected for {\it that} population.  It is possible that the four stars  
represent a distinct, distant halo substructure; evidence for additional, M giant 
halo features beyond the ``anticenter ring"
have been suggested by the analysis of Rocha-Pinto et al. (2003; see their Fig. 2).


A second ``contaminating" feature in Figure 2a is a strand of M giant velocities 
that appears to cross the main Sgr band at $\Lambda_{\sun} \sim 40-50^{\circ}$.  
Though consisting of only about a dozen stars, this strand seems to trace a similar, 
though less coherent, 
position-velocity trend as the Sgr trailing arm, and collectively these stars
suggest the possibility
of another tidal stream.  Paper I pointed out a slight
increase in the surface density of distant M giant stars in the Sgr trailing arm at 
this same $\Lambda_{\sun}$ (see Fig. 13 of Paper I), 
and offered as one explanation that the Sgr trailing arm may be
crossed by the Sgr leading arm tail at this point.  The Sgr planar distribution
of M giants gives some evidence that this crossing of the trailing and leading arms may 
exist  --- in the form of a  ``spray" of M giants to the outside of the trailing 
arm (to the lower right in Fig. 14c in Paper I).
Moreover, as shown in Law et al. (2004a,b),
the expected velocities of an extension of the leading arm 
into the Southern Hemisphere are generally consistent with the observed crossing
strand of velocities in Figure 2a.  As supporting evidence for the existence 
of this wrapped leading arm debris, the red clump stars explored by 
Majewski et al. (1999) --- which they suggest may be part of the leading Sgr arm 
wrapped around the Galactic Center --- lie at $\Lambda_{\sun}
= 27^{\circ}$ and have a mean $v_{GSR}$ consistent with this strand of stars. 
On the other hand, we cannot discount that this strand represents
an unrelated halo substructure or even a statistical fluctuation (though
the restricted $\Lambda_{\sun}$ range of this ``fluke" seems extraordinary).  
With more velocities in this region of the sky, the reality of this proposed 
halo structure can be checked.

\subsection{Nearby Southern M Giants }

A number of closer ($<13$ kpc) M giants have been observed.  Figures 2b and
2c show the distribution of velocities for these stars as a function of 
$\Lambda_{\sun}$, divided into two distance bins ($7-13$ kpc and $\le 7$ kpc).  
In both distance bins the distribution of $v_{GSR}$ with $\Lambda_{\sun}$ 
appears nonuniform; in the Southern Hemisphere ($\Lambda_{\sun} < 
165^{\circ}$) general trends of velocity with Sgr longitude are seen (though in 
opposite directions for the two bins shown).  Given the success of the more 
distant M giants to trace halo tidal features, is it possible that the nearby 
M giants are showing additional substructure closer to the Sun?  While
nothing as clear as the Sgr feature is seen in Figures 2b and 2c, nearby 
stellar streams of the same physical size as Sgr
would be expected to look less coherent in these portrayals because the 
foreshortened proximity of such streams would diversify the projections
of their bulk motion along the line of sight.  Thus, even Sgr debris
would be expected to ``smear out" with proximity to the Sun.
  
If the Figure 2a ``strand'' of crossing velocities discussed above indicates 
the presence of the Sgr leading arm crossing the trailing arm in the southern 
Galactic hemisphere, then the leading arm must pass through the Galactic plane
(on this side of the Galactic Center).  
The data in Paper I as well as models of Sgr disruption (e.g., Ibata et al. 2001; 
Law et al. 2004a,b) suggest that this may take place near the Sun.  If so, then stars with 
a large ($\gtrsim 100$ km s$^{-1}$), positive velocity should contribute to the 
nearby southern sample.  Such stars are found and, especially in the $d < 7$ 
kpc sample, well outnumber stars with large negative radial velocities.

However, these trends are complicated by the expected contribution of M giants 
from the Intermediate Population II (IPII) thick disk.  We note, for
example, that the $< 13$ kpc
sample shows an apparent velocity trend with {\it Galactic} longitude (Fig. 4)
consistent with expectations for a prograde rotating population observed 
at high latitudes (the vast majority of the $< 13$ kpc stars are
at $|b| > 45^{\circ}$).  
This signal (positive $v_{GSR}$ for $0 < l < 180^{\circ}$ and 
negative $v_{GSR}$ for $180^{\circ} < l < 360^{\circ}$) is marginally
stronger (especially in the first and second Galactic quadrants) 
for stars within 5.5 kpc of the Galactic plane than for stars 
farther than 5.5 kpc (Fig. 4), and this is what might be expected from
a contribution of stars with an IPII spatial and velocity distribution 
(e.g., Majewski 1992).  For example, 
many of the $v_{GSR} \sim +100$ km s$^{-1}$ stars in 
Figure 2c correspond to stars near $l = 90^{\circ}$ in Figure 4, the
direction of maximum rotational signal for a disk population.
Unfortunately, the tilt of the Sgr plane combined 
with the prograde Sgr orbit produces 
a similar Galactic longitude trend to that produced from simple, slow
rotation of the IPII. 
Because of the expected differences in {\it transverse velocities}
expected for IPII versus Sgr stars at the Galactic caps, proper motions 
would provide a useful discriminator;  we address this 
question in future work.  

On the other hand, noting that the majority of stars observed here have
$|b| > 45^{\circ}$, it is hard to attribute any with observed
$|v_{GSR}|$ more extreme than $220/\sqrt{2} = 155$ km s$^{-1}$ 
to any kind of disk contamination.  Even this velocity limit, 
corresponding to the 220 km s$^{-1}$ rotation speed of nearby 
thin disk stars, may be overly conservative since no such Population I 
stars are likely to find their way into our sample:
In the first place, the thin disk as a major contributor to any of the
{\it trends} discussed is ruled out 
because of both the size of the velocity variations observed as well as 
because the overall observed velocity dispersion is too high.  Recently, 
Bettoni \& Galletta (2001) obtained radial velocities for a sample of bright 
M giants near the Southern ecliptic (a plane relatively near that of the Sgr 
orbital plane), and obtained a velocity dispersion of only 34 km s$^{-1}$.  
The Bettoni \& Galletta sample, predominantly with distances $200-500$ pc 
(median distance 369 pc) and a mean $v_{hel}$ of 7 km s$^{-1}$, is 
representative of a thin disk M giant population.  
Secondly, the distances from the 
Galactic plane of the bulk of the nearby M giant sample are also much too 
large for thin disk M giants, even accounting for the possible overestimate of
the distance of these stars from assuming too low a metallicity; extrapolating  
the color-magnitude relation trends derived by Ivanov \& Borissova (2001), the mean
distance of a solar metallicity M giant at $J-K \sim 1$ is about 1.6 times closer 
than we have inferred using the mean Sgr M giant color-magnitude relation.  
Though a substantial correction, this is still too small to suggest a major 
contribution to our $< 7$ kpc sample (within which the closest star 
is 3.5 kpc distant) from thin disk M giant contaminants.
Another possibility, that the nearby samples are heavily contaminated by 
stars that are misidentified disk M {\it dwarfs}, 
would imply such an overall breakdown of the M giant/M dwarf
discrimination that is difficult to understand how Sgr could have been 
traced at all in Paper I and why the velocity distribution for the more distant
Southern Hemisphere stars (Fig. 2a) works so well (i.e., with no substantial
contribution from contaminants, even though these candidate
``more distant" stars are at magnitudes
where an even larger M dwarf contamination might be expected).  
The nearby stars are already so bright 
that if any were actually an M dwarf it would have to be extraordinarily close to the Sun.
%
%

In the end, we conclude that most or all of the high velocity (at least
$|v_{GSR}| > 150$ km s$^{-1}$, and probably even less than this limit) 
M giants shown
in Figure 2 {\it at all distances} must be from the halo, and their 
predilection for a non-uniform distribution in $\Lambda_{\sun}$ that does 
{\it not} match 
the expected rotation trend with Galactic longitude is highly suggestive 
of the presence of additional halo substructure beyond the obvious
Sgr trailing arm.  Clearly some of this substructure is in the nearby
Galaxy, and some could correspond to the Sgr leading arm passing near the Sun.
The latter is suggested by the fact that a large number of the closer, 
highest positive $v_{GSR}$ stars cluster at the South Galactic Cap 
($\Lambda \sim 75^{\circ}$), which is where one would expect to
see the highest velocities from nearby Sgr leading arm material  
egressing from the disk.

The semi-qualitative descriptions offered here for 
potential additional halo substructures within the present M giant 
radial velocity sample 
are admittedly wanting in rigor, but we have resisted 
presenting more detailed assessments at this time because future
contributions (e.g., Majewski et al. 2004b)
will offer substantially enlarged radial velocity samples as a 
result of our ongoing spectroscopic program, and because numerous
analyses we have attempted on the present database are no more convincing
than the arguments presented here as a result of the irregular sky 
sampling among the closer stars.
With a more systematic sampling of both the Northern and Southern Hemisphere
samples we hope to clarify these tentative and tantalizing features.

\subsection{Northern M Giants }

The sampling of Northern Hemisphere M giants observed with the Swope Telescope 
is meager.  The nearby ($d < 13$ kpc) M giant sample has a mean velocity
near $-8$ km s$^{-1}$ and a dispersion of 93 km s$^{-1}$.  

Four more distant ($33-41$ kpc) stars in the far side of the leading tail
have been observed so far.  As we show in future papers, these have
velocities consistent with membership in the distant Northern Loop of 
leading arm Sgr debris, which is roughly traced out by their velocity
trend on the left side of Figure 2a.

\section{The Coldness of the Sgr Trailing Stream and Limits on Halo Lumpiness
and the Sgr Core Mass}

Current Cold Dark Matter (CDM) models for the formation of structure in the 
universe predict that MW-like galaxies should contain substantial halo substructure
at current epochs as a result of the accretion of thousands of subhalos over a Hubble time
(e.g., Navarro et al.\ 1996, 1997).  Because the MW currently has only eleven 
known {\it luminous} satellite galaxies, it is commonly held that the bulk of the 
subhalos must be made up of pure dark matter (see Klypin et al. 1999,
Moore et al. 1999).
If so, these dark matter lumps should make their presence known
through the heating of dynamically cold, {\it luminous} stellar systems
(Moore et al. 1999, Font et al. 2001; see also Moore 1993).  
The implied coldness of tidal streams, in principle, should make them
particularly sensitive to substructure lumpiness
in the Galactic halo (Moore et al. 1999, Murali \& Dubinski 1999, 
Johnston, Spergel \& Haydn 2002, 
Ibata et al. 2001, 2002, Mayer et al. 2002).  However, as we noted above
(\S 3.1.1), the velocity dispersion of a stream is also a function of
the mass of the stream progenitor (see also Law et al. 2004a,b).  
Thus, the present
velocity dispersion of the Sgr trailing tail reflects contributions from both 
the initial dispersion of tidally released debris and subsequent 
heating of that debris imparted by encounters with large halo masses.   
Attributing all of the dispersion to one or the other of these phenomena provides upper 
limits to the effects of each.

In Law et al. (2004a,b) we find that the velocity dispersion data of the trailing arm presented
here is best fit by a smooth Milky Way halo model
wherein the present bound Sgr core has a mass of 3.0 $\times 10^{8} M_{\sun}$, 
corresponding to $(M/L)_{V, tot} = 21$ in solar units.\footnote{This specific 
limit corresponds to the specific core {\it structure} adopted in Law et al. (2004a,b), 
and may vary somewhat depending on reconfigurations of the mass distribution in the
core.  See discussion in Law et al. (2004a,b).} 
However, should the 
velocity dispersion of the tidal debris have any contribution from heating by 
lumps in the Galactic halo, the upper limit on the mass of the Sgr core
is reduced.
That the velocity dispersion appears to increase as a function 
of $\Lambda_{\sun}$ (see \S3.1.1) could be a signature of the cumulative effects 
of 
such heating, though other possible explanations for the trend were 
enumerated earlier (\S3.1.1).  Nevertheless, because of the upper limits
on the heating it provides, it is a worthwhile exercise to proceed under
the assumption that the spatial and velocity dispersion of the stream
is entirely from scattering effects of Galactic subhalos.  Because of the improvement in the
tracing of the Sgr stream in Paper I and in deriving its dynamics with the 
velocities here, we can reassess the Sgr stream as a delimiter of cosmological models,
which predict different degrees of halo lumpiness.
 
Johnston, Spergel \& Haydn (2002) give a prescription for a ``scattering index" 
for tidal debris that is sensitive to halo lumpiness, and then applied it to the 
Sgr stream as traced by the carbon stars of Totten \& Irwin (1998).  This 
index measures perturbations in the positions and velocities of tidal debris stars
induced by lumps in the halo under the assumption of an initial tidal debris population
with no velocity dispersion --- thus the index provides an 
upper limit to the scattering for realistic debris.  
The variables $d\theta$ and $\psi$ in equations 1 and 2 of Johnston et al. (2002) 
correspond to $\beta$ and $\Lambda_{\sun}$ for Sgr debris in Paper I.  

We use both the 
$\Lambda_{\sun} = 25^{\circ}-90^{\circ}$ and $\Lambda_{\sun} = 25^{\circ}-150^{\circ}$ samples
from \S3.1.1 to derive the $B_m$ and $B$ parameters from Johnston et al. (see Fig. 5).
The M giants in the latter sample are limited
to the past several orbits of drift from the parent satellite, so would measure
the degree of heating for about 2.5 Gyr or so of dynamical age.
The net result from either sample yields $B_m$ values smaller than
Johnston et al. (2002) obtained for the assumed Sgr carbon stars.  For the smaller
longitude range, a net $B=0.031$ is obtained, leading to 
a more restrictive limit on lumpiness than the $B=0.037$ obtained for the
carbon stars (though also corresponding also to a more restricted length
of the tidal tail, and less than the past $\sim2$ Gyr of heating).  
This smaller $B$ value is consistent with the Johnston et al. simulations
of heating in a smooth halo containing just one LMC-like (mass and orbit) lump;  
however, at least some realizations of lumpier halos in the Johnston et al. analysis are
also consistent with the degree of scattering identified here.  Unfortunately,
this reflects the vagary of this type of halo probe: Dynamical heating in CDM halos 
tends to be dominated by the most massive lumps.  
Nevertheless, there may yet remain discriminatory power in this test of lump 
heating if the scattering statistic can be lowered yet more after removal of
the contribution to the scattering by the zero-age dispersion of the debris stream.
The required detailed study of velocities for stars around the Sgr tidal radius 
should be available soon.  However, it is worth noting that the $B$ value for 
the entire $\Lambda_{\sun} = 25^{\circ}-145^{\circ}$ M giant sample rises to 0.036,
which could be a reflection of the increased effects of halo heating over the 
longer length/time scale of the debris represented in that stellar sample. 


To some extent, the present result of consistency with a relatively smooth halo 
was already implied by the demonstration in Paper I and earlier by Ibata et al. (2001) 
that the Sgr debris shows minimal orbital {\it precession},  
which is another effect induced by non-spherical potentials.  Lumpy halos
would impart some degree of precession, according to models by
Mayer et al. (2002).

\section{Globular Cluster Comparison to Trailing Arm Velocities}

Since globular clusters were immediately recognized to be a feature of the
the Sgr core (Ibata et al. 1995, 1997), a
number of other globular clusters have been nominated as potential
members of the extended Sagittarius debris stream (Lynden-Bell
\& Lynden-Bell 1995, Irwin 1999, Dinescu et al. 1999, 2000, Palma et al. 2002,
Bellazzini et al. 2002, 2003a).  Obviously, a feature of such clusters is
that they will share both positional and velocity coincidence with the
stellar tidal arms, and various methods have been employed among these
earlier studies to determine whether this holds true on a cluster by 
cluster basis.  However, these earlier studies were hampered by a poorly
defined orbit for Sgr due to incomplete knowledge of the position and velocity  
of either the core (for which the proper motion is only roughly determined; see, 
e.g., Ibata et al. 1997)
or its debris.  In their recent analysis, Bellazzini et al. (2003a) compared
the positions of clusters with the best existing model orbit for Sgr, and
presented a list of potentially interesting associations. 
However, now that much improved data on the Sgr debris stream
are being assembled, the problem can be revisited with a direct
{\it empirical} comparison to both velocities and positions.

In a future contribution we will conduct a more thorough and extensive analysis
with our larger set of radial velocity data in both hemispheres as well as our Sgr debris models.  
However, given the data presented in Figure 2, a number of immediate checks on
previously postulated as well as new Sgr cluster candidates can be made
against the well-defined trailing arm velocities in the Southern
Hemisphere.  In Figure 6 we present the same M giant data as in Figure 2a, 
but include the globular
clusters in the indicated Sgr longitude range with $d>13$ kpc
and lying within 10 kpc of the nominal Sgr plane from Paper I.  In that
latter criterion, we have adopted a liberal breadth of the stream, 
i.e. twice the lower limit from \S3.1.3.  

As might be expected from previous work showing their association with the
Sgr system, the four Sgr core clusters M54, Terzan 7, Terzan 8 and Arp 2 have velocities
consistent with an extrapolation of the M giant velocity trend.  At the
other end of the M giant swath, we see that the cluster Pal 2 is similarly
situated along an extrapolation of the trailing arm velocities towards the 
Galactic anticenter, if one takes into account the apparent turn-up
in the velocities at $\Lambda_{\sun} \sim 160^{\circ}$.  
Lynden-Bell \& Lynden-Bell (1995) first suggested
a potential association of this cluster with Sgr.  Bellazzini et al. (2003a)
also demonstrate a nice velocity match of this cluster to their Sgr orbit model, 
but the cluster lay almost 15 kpc in the orbital plane from the Sgr
orbital path (see their Table 1).  However, a comparison of Pal 2's 
position with the position of Sgr
{\it debris} at the same longitude finds much better agreement: its distance of
27.6 kpc places Pal 2 just on the inner edge of M giant debris at this
longitude ($\Lambda_{\sun}=153.8^{\circ}$ (see Fig. 10 of Paper I).  Moreover,
Pal 2 is 5.95 kpc from the Sgr mid-plane, only 1 kpc larger than the 
width found above (\S3.1.3).  From the present evidence,
including the new velocity evidence presented in Figure 6, 
Pal 2 seems a reasonably good candidate for a globular cluster member of the Sgr 
trailing arm debris.  Based on our models (Law et al. 2004a,b), the position and
velocity of Pal 2 is most consistent with it being among the debris torn from
Sgr two perigalactica ago ($\sim 1.5$ Gyr ago), although the overlap of debris with
different energies at the same orbital phase along the trialing arm means that
we cannot exclude an escape of the cluster from Sgr one to four periGalactica
ago ($\sim 0.75-3.0$ Gyr ago).

From a derived orbit based on a new proper motion measurement of the cluster,
Pal 12 has already been discussed as a Sgr cluster by Dinescu et al. (2000).
In Paper I (see Fig. 17 of that paper) it was shown that Pal 12 has a distance
that places it right amidst the trailing arm debris.  
Subsequently, both Martinez-Delagado et al. (2002)
and Bellazzini et al. (2003b) have shown that Sgr stellar debris lies around
the cluster based on color-magnitude diagrams of the cluster field.  
Pal 12 lies only 3.24 kpc from the nominal Sgr plane from Paper I.  
Nevertheless, based on the orbit they calculated for Pal 12
from its space velocity, Dinescu et al. (2000) argued that the cluster
is more consistent with association to {\it leading arm} Sgr debris.  Perhaps then
it is interesting that the radial velocity of Pal 12, while similar to the
M giant Sgr trailing arm velocity trend, is in fact 16.4 km s$^{-1}$ 
off the mean of the M giant trend from \S3.3.1, which is
at least 1.5 times the intrinsic dispersion of the M giants.  In addition, 
Pal 12 could plausibly lie among the more widely dispersed ``strand" of
stars proposed in \S3.2 to be
Sgr leading arm material crossing the trailing arm velocity trend.  From a comparison
to the present M giant radial velocity data alone it is difficult to discriminate 
between the leading and trailing arm possibilities, and the answer to the problem 
must rely on full space velocity comparisons.  Thus in the end we defer to the Dinescu
et al. (2000) conclusion of a Sgr leading arm association, and look forward to
both proper motion data for the M giant stream as well as an improved proper
motion for Pal 12 to settle the matter. Nevertheless, the radial velocity
data here do add weight to the conclusion that Pal 12 is a Sgr globular cluster.

The cluster NGC 6864 presents a more conflicting picture.  It lies only
3.7 kpc from the nominal Sgr plane and its
18.8 kpc distance at $\Lambda = 15.1^{\circ}$ also places it nicely in the
trailing arm of Sgr debris (see Fig. 10 of Paper I), but NGC 6864's radial velocity 
(Fig. 6) surely argues against this association.  The cluster's rather large
{\it approaching} velocity makes it hard to attribute NGC 6864 to either
trailing or wrapped leading arm material in the Southern Hemisphere. 
Although improved models or data may show a way for 
Sgr debris to have such a large negative velocity at this position in the
Galaxy, for now we must conclude that NGC 6864's association with Sgr seems unlikely.
Similar arguments hold for NGC 7492 and NGC 6981:  Although only 4.1 and 6.0
kpc from the Sgr mid-plane, respectively, and both
reasonably placed in distance (NGC 7492: $d=25.8$ kpc at $\Lambda=58.9^{\circ}$, 
NGC 6981:  $d=17.0$ kpc at $\Lambda=27.0^{\circ}$) 
near the trailing arm M giants, these two clusters both have velocities 
extremely discordant with that association, or with expectations for 
leading arm debris passing into the Southern Hemisphere.  

In conclusion, our new velocity data, when combined with the positional data for
M giants in Paper I, support the association of the clusters Pal 2 and Pal 12
with Sgr debris, but appear inconsistent with the association of NGC 6864, 
NGC 6981 and NGC 7492 to Sgr debris.

\section{Discussion}

Radial velocities have been obtained for nearly three hundred 
M giant candidates identified
in the 2MASS survey to be near the Sgr orbital plane.  The survey data here 
concentrates mainly on stars in the Southern Hemisphere.  

The radial velocity trend of the Sgr trailing arm is clear and distinct
and provides an important constraint on models of the Sgr debris stream
(see Law et al. 2004a,b).  The velocity dispersion of the trailing debris arm 
limits both the degree of lumpiness of the MW halo as well as the mass
of the Sgr bound core.
Because these two contributors to the velocity dispersion
play off one another, it may be difficult simultaneously to have 
both a large $M/L$ for the Sgr core {\it and} a lumpy halo:  At present
we limit the $(M/L)_{V,tot}$ of the Sgr core to 21 in solar units 
(see Law et al. 2004a,b), and the
heating of the Sgr stream to be consistent with expectations 
for debris encounters with only the LMC.  More stringent constraints on both may 
be possible after dynamical study of the Sgr tidal boundary to determine
the zero-age velocity dispersion of Sgr debris.

At least one other debris stream may 
be present in the distant M giant sample, and its velocity trend is consistent
with model (Law et al. 2004a,b) predictions for wrapped, leading arm material
in the Southern Hemisphere, but it may also be an unrelated
tidal stream.  The globular cluster Pal 12 may be associated
with the stars in this feature, as previously postulated by Dinescu et al. (2000), 
but we note that the radial velocity of this cluster lies within 1.5-2$\sigma$ 
of the mean for Sgr trailing arm M giants at this same location in the Galaxy.
Based on its radial velocity and positional match with Sgr M giants, we find 
compelling evidence that the globular cluster Pal 2 is associated with
Sgr trailing arm debris.

Among the closer ($< 13$ kpc) M giant 
stars observed are more than half a dozen with the very high positive radial velocities 
expected for Sgr debris in the neighborhood of the Sun, and we point to
the presence of many high velocity ($|v_{GSR}| > 150$ km s$^{-1}$), 
high latitude M giants that are difficult to account for as other than
halo substructure relatively close to the Sun.  

The data presented here are better understood within the Sgr debris
context by reference to interpretive models.  Such models are given
in Law et al. (2004a,b), in which the data presented here provide
useful observational constraints.  Moreover, additional velocities obtained in
the Northern Hemisphere will help clarify the interpretations offered here.
Future papers will present new data that increase by more than five times the present 
M giant velocity sample.  

We thank Carnegie Observatories Director A. Oemler 
for generous access to the Swope telescope and all of the
Las Campanas staff for their diligent and generous support at the telescope.
The results presented in this publication make use of data from
the Two Micron All Sky Survey (2MASS), which is a joint project of the
University of Massachusetts and the Infrared Processing and Analysis
Center (IPAC), funded by the National Aeronautics and Space Administration
and the National Science Foundation.  This work was supported by
National Science Foundation grant AST-0307851 and a
Space Interferometry Mission Key Project grant, NASA/JPL contract 1228235.
MFS acknowledges support from NASA/JPL contract 1234021.  MDW was supported
in part by NSF grant AST-9988146.  This work was also partially supported
by the Celerity Foundation.  We appreciate comments from the referee that 
have improved the clarity of the paper.

\begin{figure}
\plotfiddle{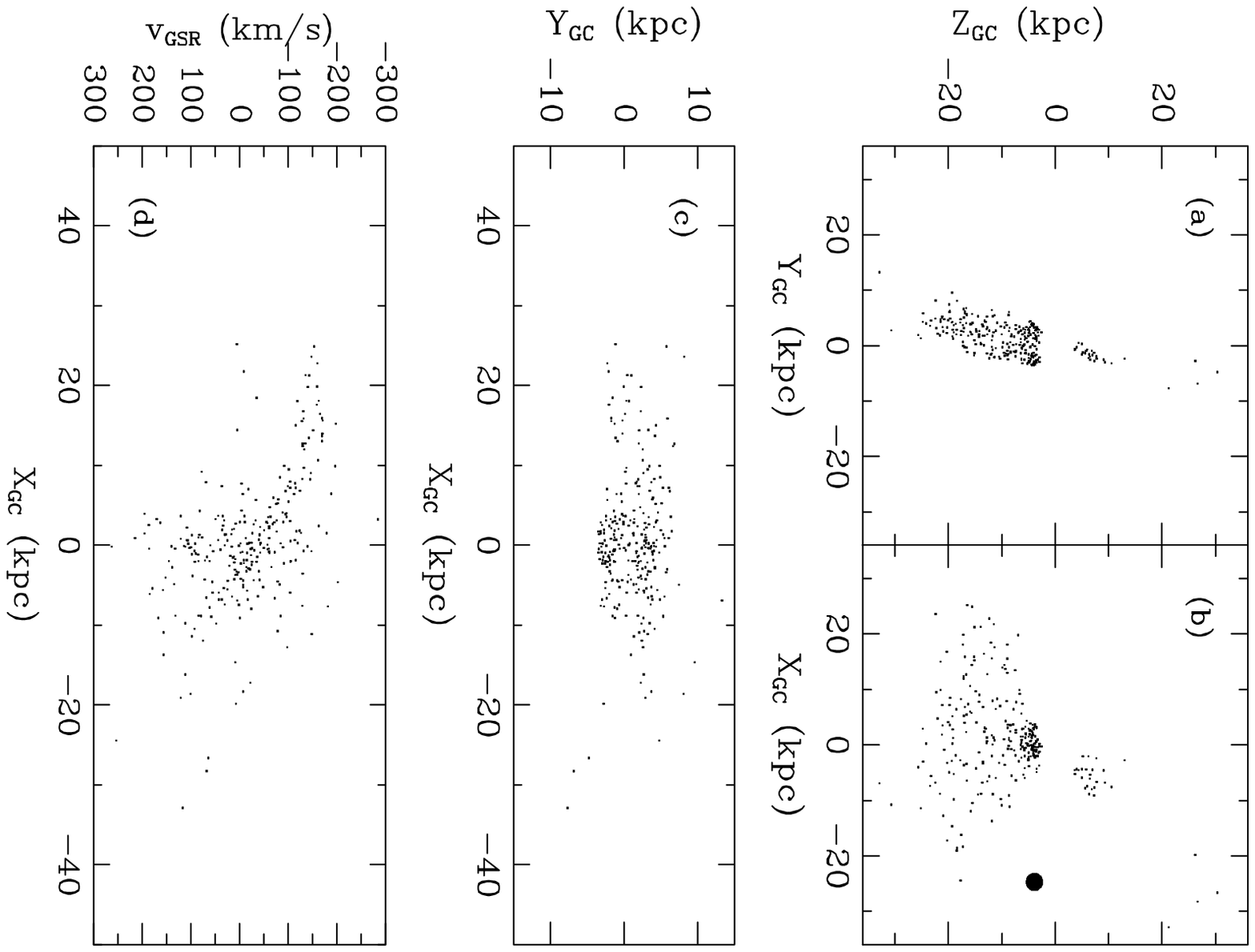}{5.5in}{90}{90}{90}{370}{-65}
\plotfiddle{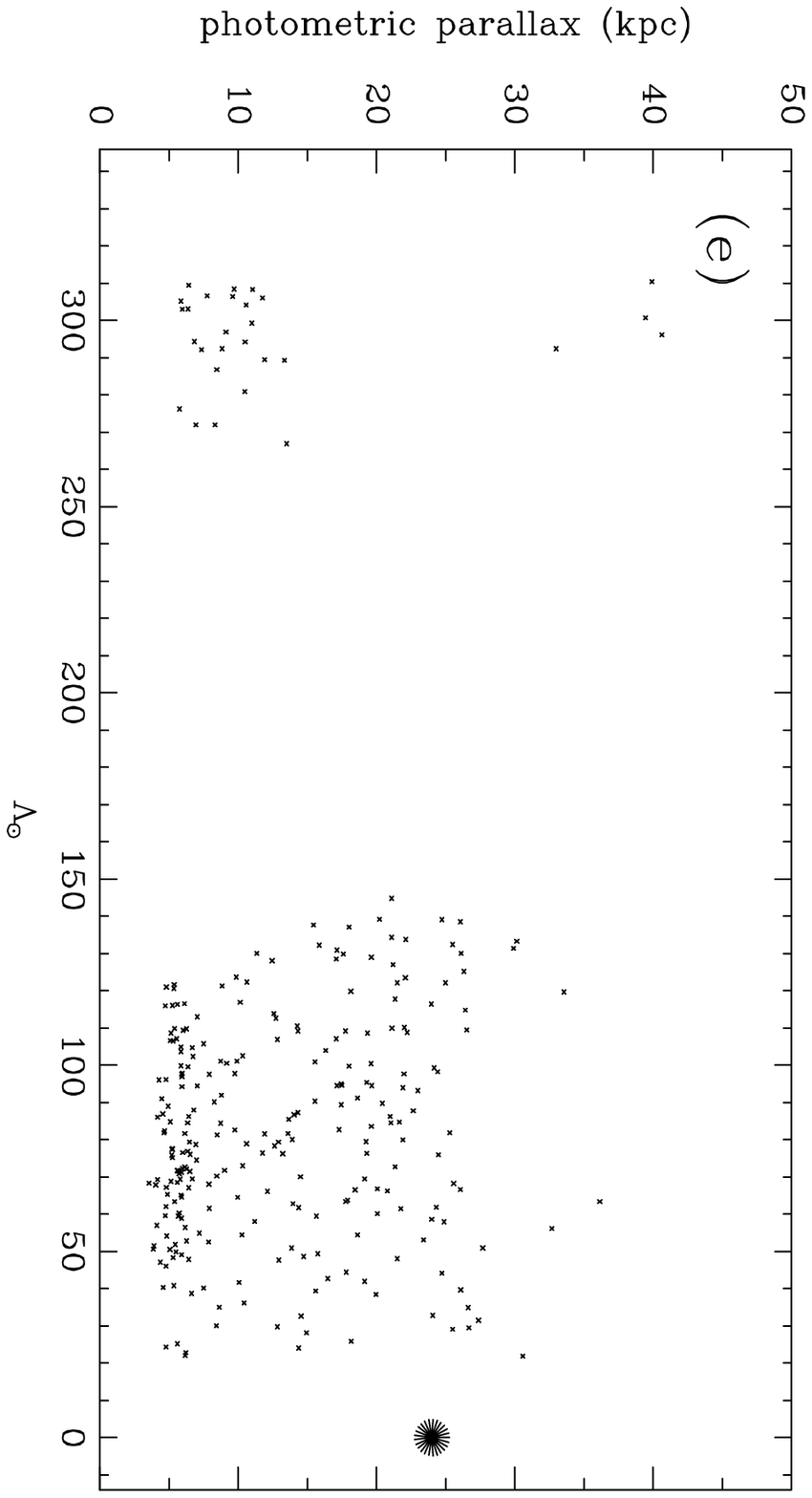}{1.75in}{90}{60}{60}{250}{-150}
\caption{
Spatial distribution of M giant candidates for which radial velocities have been 
obtained.  Compare this figure to the full distribution of M giants shown in
Figures 7 and 11 of Paper I.
In the bottom panel, the distribution of M giant candidate photometric parallax
distances is given as a function of Sgr
$\Lambda_{\sun}$ longitude as defined in Paper I.  Compare this distribution
to Figure 10 in that paper.  The large dot corresponds to the location of the
Sgr core.
  }
\end{figure}

\begin{figure}
\plotfiddle{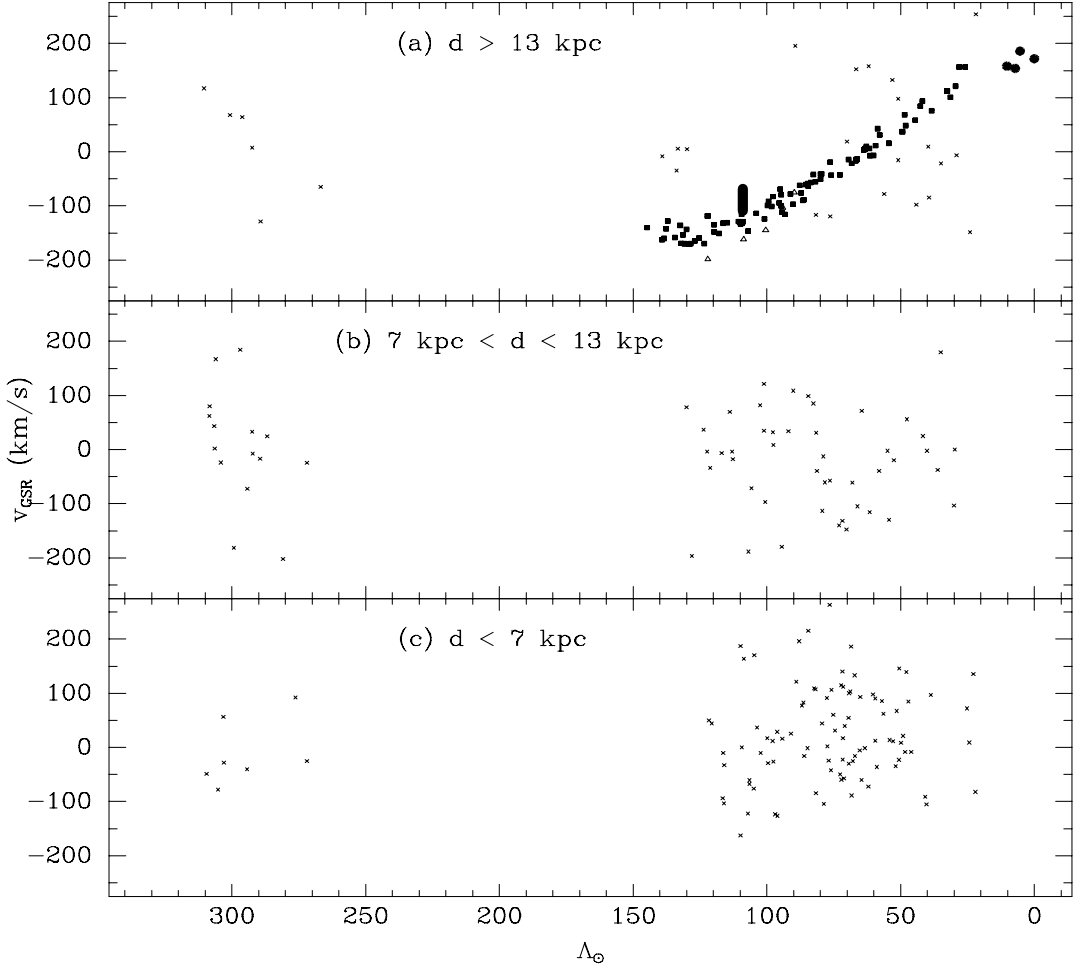}{5.5in}{90}{80}{80}{320}{-65}
\caption{
Galactic standard of rest velocities 
of M giants as a function of Sgr orbital longitude: (a) M giants with 
large ($>13$ kpc) photometric parallax distances.  The larger circles 
represent the velocities of the Sgr core globular clusters.  Measurements of the Sgr 
core velocities by Ibata et al. (1997) are consistent with, and obscured by, the cluster 
points at the same $\Lambda_{\sun}$.  The elongated, vertical mark at 
$\Lambda_{\sun}=110^{\circ}$ shows the position of the
Y03 measurement, where the length represents the $\pm$ dispersion found by those authors.
In this panel, we show the stars used in the parametric fits of \S3.1.1 as
solid squares, and stars rejected from those fits as crosses.  Stars
observed on the night of UT 2002 September 20, which may have large (systematic
and random) errors, 
were not used in those fits; those stars are shown in this panel with open triangles.
(b) Velocities of M giant stars at intermediate distances.  
(c) Velocities of M giants with nearby ($\le 7$ kpc) photometric parallax distances.  
  }
\end{figure}

\begin{figure}
\plotfiddle{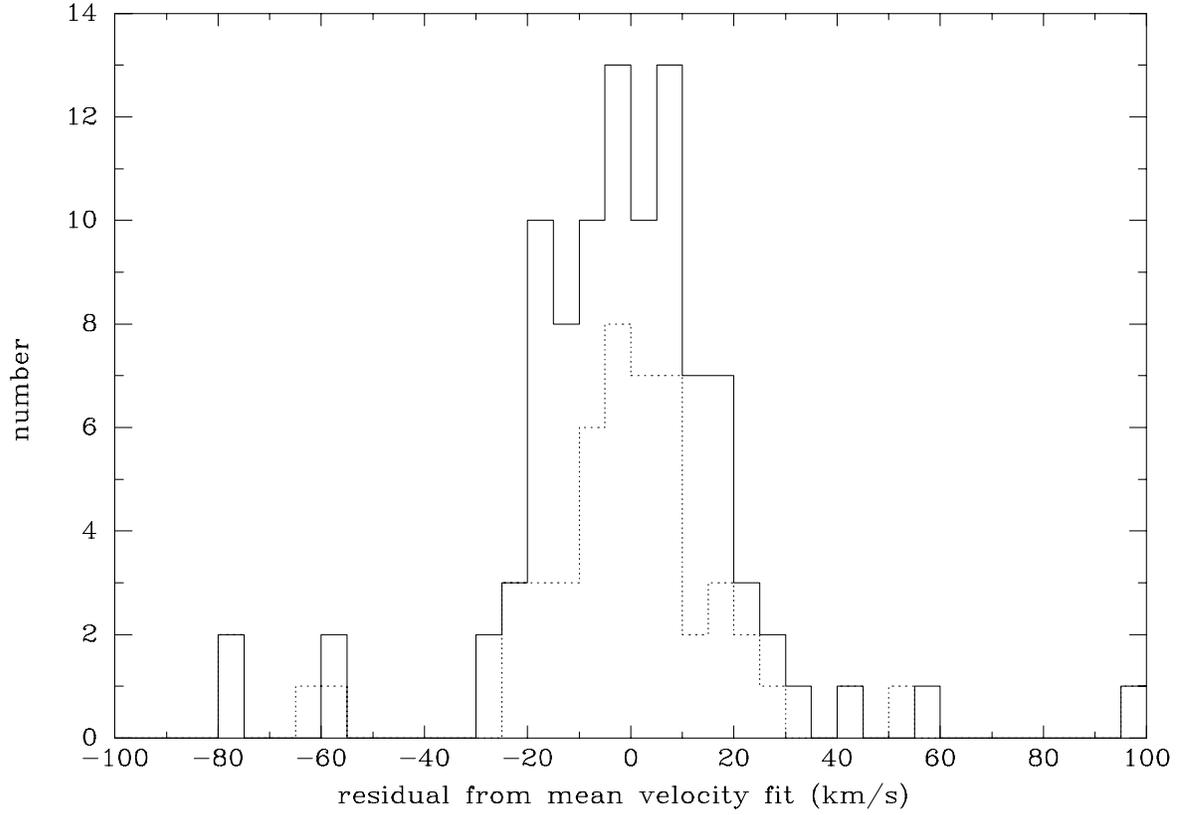}{5.5in}{90}{80}{80}{300}{-5}
\caption{
The distribution of Sgr trailing stream radial velocities about the mean third order
fits described in \S3.1.1 as a function of $\Lambda_{\sun}$.  
The {\it solid line} shows the dispersion
for $25^{\circ} < \Lambda_{\sun} < 150^{\circ}$, and the {\it dotted line} is the
dispersion for $25^{\circ} < \Lambda_{\sun}<  90^{\circ}$. 
 }
\end{figure}

\begin{figure}
\plotfiddle{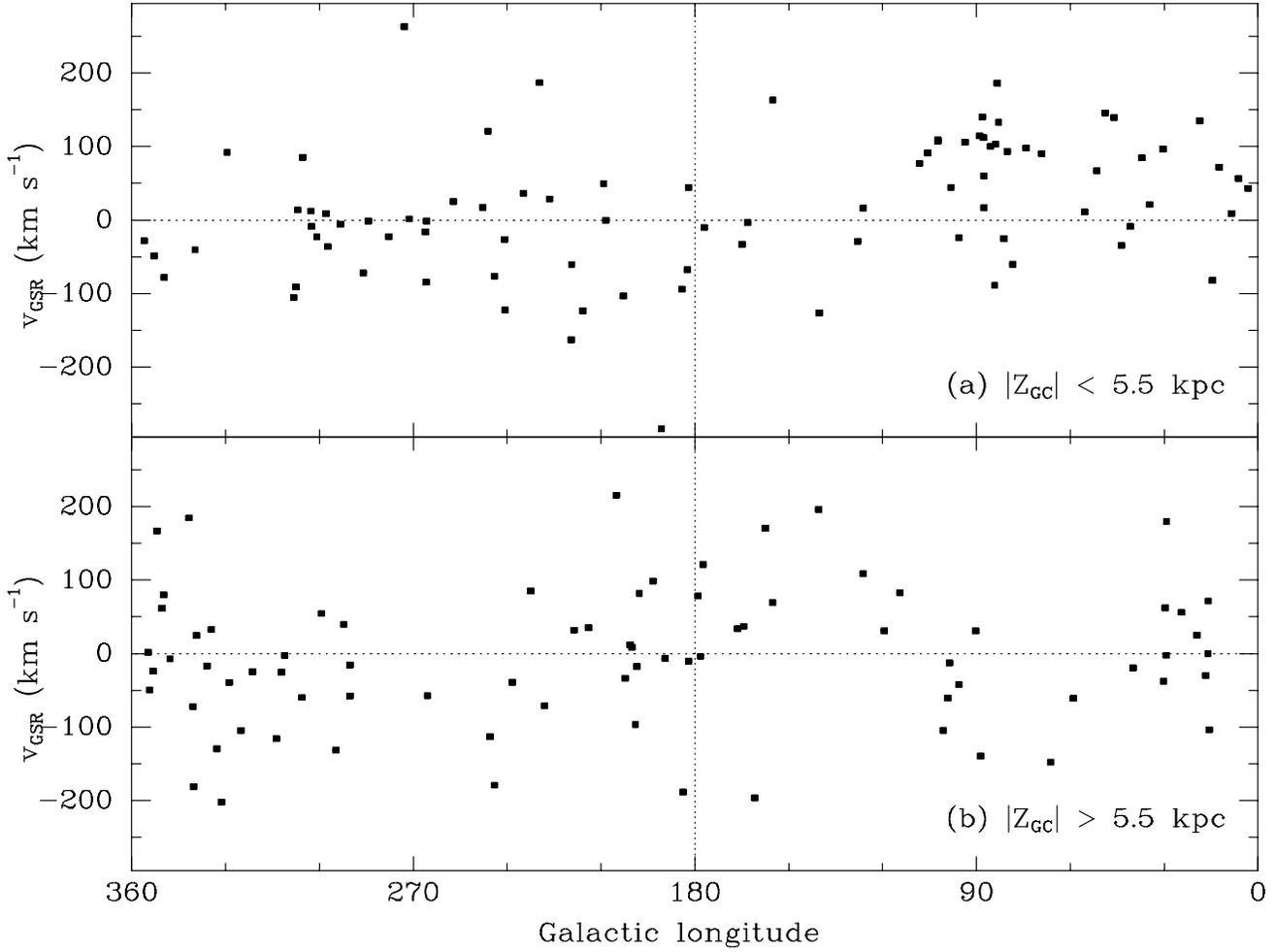}{5.5in}{90}{90}{90}{310}{-5}
\caption{
The distribution of $d < 13$ kpc M giant velocities as a 
function of Galactic longitude.  (a) The stars with $Z_{GC} < 5.5$
kpc.  (b) Stars with $Z_{GC} > 5.5$.  A division at $Z_{GC} = 5.5$ 
kpc corresponds to the extent of the IPII as observed by Majewski
(1992) and also roughly divides the $d < 13$ kpc sample in half. 
 }
 
\end{figure}

\begin{figure}
\plotfiddle{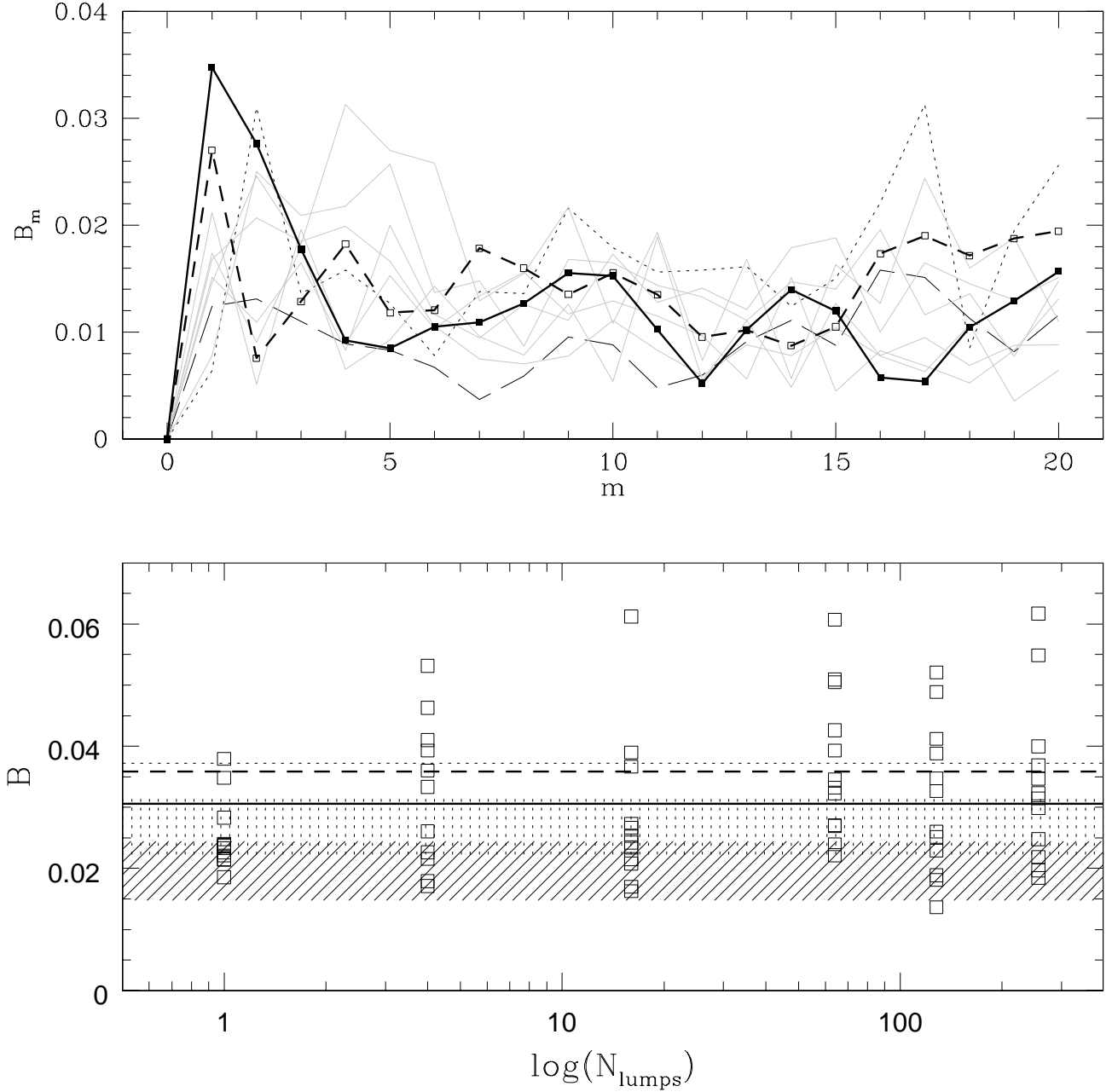}{6.05in}{0}{90}{90}{-265}{-150}
\caption{Figure adapted from Johnston, Spergel \& Haydn (2002)
showing (top) the Fourier series components, $B_m$, of their 
scattering index, and (bottom) the total scattering index, $B$,
for both smooth and lumpy halo models.  In the top panel, the thin solid lines 
show the results for various halo simulations with 128 halo lumps (note,
Johnston et al. erroneously state that these lines represent 
$N_{lump} = 256$), while the long-dash line represents
the Fourier series components for a smooth halo.  The dotted line
reproduces the carbon star results given in Johnston et al.  
Results for M giants in the range 
$25^{\circ} < \Lambda_{\sun} < 90^{\circ}$
are shown as the thick solid line, while M giants in the range
$25^{\circ} < \Lambda_{\sun} < 145^{\circ}$ are represented by the
thick dashed line.  The same M giant line types are shown in the 
lower panel.  The diagonally hashed region 
in the bottom panel shows the range expected for debris
evolution in a smooth potential while the upper shading is
the range of expected results for Sgr debris simulations 
having a halo contianing only one lump of mass $10^{10} M_{\sun}$ 
on an LMC-like orbit.  
The open squares in the bottom panel 
are the scattering indices obtained from simulations of halos with 
different numbers of lumps and random relative orientations of lump 
and debris.  The centroid of the open squares can be interpreted as 
a ``median case" for a given $N_{lumps}$.
See Johnston et al. for further explanation of this figure.
}
 
\end{figure}

\begin{figure}
\plotfiddle{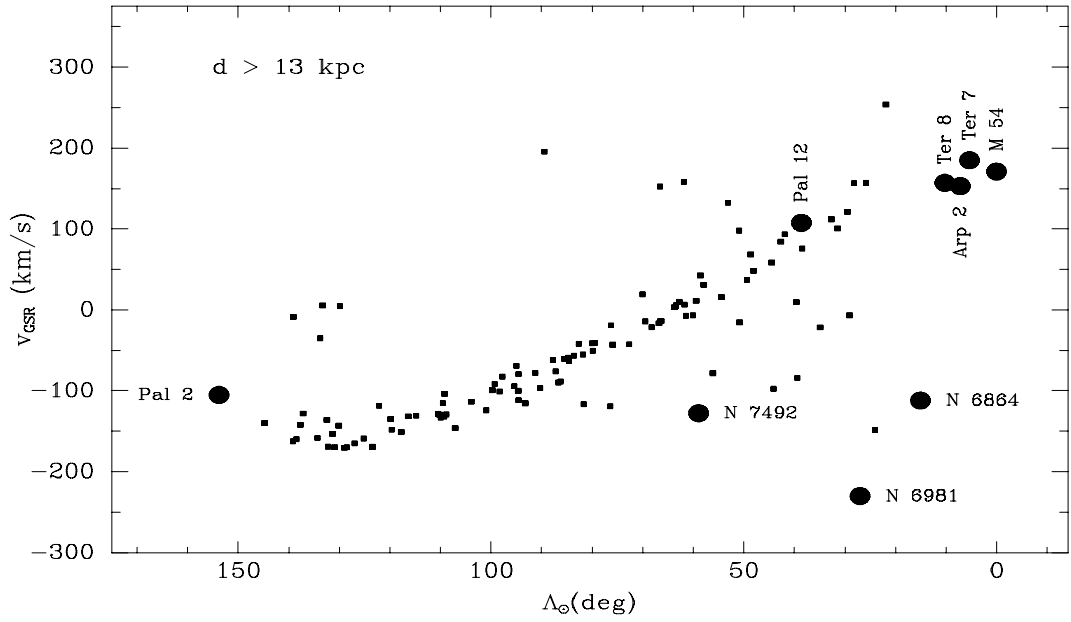}{6.15in}{0}{110}{110}{-315}{-250}
\caption{Zoomed in region of Figure 2a, comparing the M giant
velocity trend with Southern Hemisphere globular clusters 
also with $d>13$ and lying within 10 kpc of the nominal Sgr plane.}
\end{figure}


\begin{deluxetable}{lrrrr}
\tabletypesize{\footnotesize}
\tablewidth{0pt}
\tablecaption{Scatter in Radial Velocity Measures, by Observing Run}
\tablehead{
&\multicolumn{2}{c}{Bright RV Stds.}&\multicolumn{2}{c}{2MASS M giants\tablenotemark{a}}\\
\colhead{UT Dates}&     \colhead{Scatter (km s$^{-1}$)}& \colhead{No. Stars} &  \colhead{Scatter
(km s$^{-1}$)}&  \colhead{No. Stars}   }
\startdata

15 -- 19 Jul 2002        &  4.4   & 14 & 4.6 & 1            \\
29 -- 30 Jul 2002        &  8.2   & 21 & 7.4 & 5           \\
31 Jul -- 01 Aug 2002\tablenotemark{b} &  4.2   &  9 & 4.2 & 2 \\
19 -- 20 Sep 2002\tablenotemark{c}        & $>20$\ \ \   & & \nodata& 0 \\
23 -- 30 Sep 2002        &  5.7   & 75 & 6.6 & 7    \\
21 -- 26 Nov 2002        &  8.8   & 21 & 0.6 & 1   \\

\enddata
\tablenotetext{a}{From multiple measures of stars on a single run. See Table 2.}
\tablenotetext{b}{The spectrograph setup focused on the Calcium IR triplet region for this run.}
\tablenotetext{c}{There were problems with the slit width on this run. Only 12 stars from this run
contributed to the sample.}
\end{deluxetable}

\begin{deluxetable}{rrrrrrrl}
\tabletypesize{\footnotesize}
\tablewidth{0pt}
\tablecaption{M Giants with Repeat Observations}
\tablehead{
\colhead{NAME}&     \colhead{ccd\#}& \colhead{$v_{hel}$} &  \colhead{$v_{GSR}$}&  \colhead{$CCP$} & \colhead{$Q$} & 
\colhead{$\sigma{(RV)}$ }&\colhead{RV date} }
\startdata

$2326237-250037$\tablenotemark{a} & \nodata      &$\mathbf{-27.7}$&$   11.2$ & 0.81&  7&        4.6                           & 2002-07-16/18  \\
 \nodata                   & 2055  &$   -30.9$&$    8.0$ & 0.79&  7& \nodata                    & 2002-07-16  \\
 \nodata                   & 4058  &$   -24.4$&$   14.4$ & 0.83&  7& \nodata                    & 2002-07-18  \\
&&&&&&&\\
$1515571-074734$\tablenotemark{b} & \nodata      &$\mathbf{   -14.5}$&$  -23.9$ & 0.37&  7&  7.6                  & 2002-07-29  \\
 \nodata                   & 5014  &$    -9.1$&$  -18.5$ & 0.36&  5& \nodata                    & 2002-07-29  \\
 \nodata                   & 5015  &$   -19.9$&$  -29.3$ & 0.38&  5& \nodata                    & 2002-07-29  \\

$1519212-100704$\tablenotemark{a} & \nodata      &$\mathbf{   181.2}$&$  167.0$ & 0.67&  7&    8.7                & 2002-07-29  \\
 \nodata                   & 5007  &$   175.0$&$  160.8$ & 0.75&  7& \nodata                    & 2002-07-29  \\
 \nodata                   & 5008  &$   187.3$&$  173.1$ & 0.58&  7& \nodata                    & 2002-07-29  \\

$1522364-132713$\tablenotemark{a} & \nodata      &$\mathbf{   101.9}$&$   79.9$ & 0.68&  7&   12.9                & 2002-07-29  \\
\nodata                    & 5010  &$    92.8$&$   70.8$ & 0.77&  7& \nodata                    & 2002-07-29  \\
\nodata                    & 5011  &$   111.0$&$   89.0$ & 0.59&  7& \nodata                    & 2002-07-29  \\

$1538463-023544$\tablenotemark{a} & \nodata      &$\mathbf{    21.6}$&$   43.2$ & 0.72&  7&     2.5               & 2002-07-29  \\
\nodata                    & 5019  &$    19.8$&$   41.4$ & 0.71&  7& \nodata                    & 2002-07-29  \\
\nodata                    & 5020  &$    23.4$&$   45.0$ & 0.72&  7& \nodata                    & 2002-07-29  \\

$1441233-124407$\tablenotemark{c} & \nodata      &$\mathbf{  -133.9}$&$ -181.1$ & 0.57&  7&     5.1               & 2002-07-30  \\
\nodata                    & 6031  &$  -139.8$&$ -187.0$ & 0.48&  7& \nodata                    & 2002-07-30  \\
\nodata                    & 6032  &$  -131.0$&$ -178.1$ & 0.62&  7& \nodata                    & 2002-07-30  \\

&&&&&&&\\
$2050020-345336$\tablenotemark{a}~\tablenotemark{d} & \nodata      &$\mathbf{   -20.5}$&$    9.0$ & 0.79&  7&   4.6               & 2002-07-30/31  \\
\nodata                    & 6052  &$   -23.7$&$    5.7$ & 0.83&  7& \nodata                    & 2002-07-30  \\
\nodata                    & 7018  &$   -17.2$&$   12.2$ & 0.74&  7& \nodata                    & 2002-07-31  \\

$1429162-075540$\tablenotemark{a}~\tablenotemark{d} & \nodata      &$\mathbf{   -30.9}$&$  -72.1$ & 0.69&  7&   6.0               & 2002-07-30/31  \\
\nodata                    & 7010  &$   -26.6$&$  -67.8$ & 0.67&  7& \nodata                    & 2002-07-31  \\
\nodata                    & 6027  &$   -35.1$&$  -76.3$ & 0.70&  7& \nodata                    & 2002-07-30  \\

&&&&&&&\\
$1412322-052131$\tablenotemark{a} & \nodata      &$\mathbf{   -84.2}$&$ -129.0$ & 0.77&  7&     0.6               & 2002-07-31  \\
\nodata                    & 7006  &$   -84.6$&$ -129.4$ & 0.71&  7& \nodata                    & 2002-07-31  \\
\nodata                    & 7008  &$   -83.7$&$ -128.6$ & 0.83&  7& \nodata                    & 2002-07-31  \\

$0013189-230153$\tablenotemark{c} & \nodata      &$\mathbf{    -6.5}$&$   19.2$ & 0.46&  7&     7.7               & 2002-07-31  \\
\nodata                    & 7020  &$   -15.5$&$   10.2$ & 0.35&  4& \nodata                    & 2002-07-31  \\
\nodata                    & 7021  &$    -2.0$&$   23.7$ & 0.52&  6& \nodata                    & 2002-07-31  \\

&&&&&&&\\

$0202225+003225$\tablenotemark{b} &\nodata       &$\mathbf{   133.1}$&$  170.5$ & 0.46&  7&  6.1                  & 2002-09-26  \\
\nodata                    & 5115  &$   137.4$&$  174.8$ & 0.43&  7& \nodata                      & 2002-09-26  \\
\nodata                    & 5117  &$   128.8$&$  166.1$ & 0.48&  7& \nodata                      & 2002-09-26  \\

$0332342-252639$\tablenotemark{c} &\nodata       &$\mathbf{   -66.2}$&$ -162.8$ & 0.69&  7&  3.5          & 2002-09-26  \\
\nodata                    & 5151  &$   -70.2$&$ -166.9$ & 0.35&  6& \nodata                      & 2002-09-26  \\
 \nodata                   & 5153  &$   -64.2$&$ -160.8$ & 0.86&  7& \nodata                      & 2002-09-26  \\
\tablebreak\\

$2321309+082002$\tablenotemark{c} &\nodata       &$\mathbf{    -8.8}$&$  140.2$ & 0.63&  7& 2.8             & 2002-09-26  \\
\nodata                    & 5086  &$   -12.0$&$  137.0$ & 0.43&  7& \nodata                      & 2002-09-26  \\
\nodata                    & 5088  &$    -7.2$&$  141.8$ & 0.73&  7& \nodata                      & 2002-09-26  \\

$2350361-200216$\tablenotemark{c} & \nodata      &$\mathbf{   -61.1}$&$  -14.2$ & 0.42&  7&     5.1               & 2002-09-27  \\
\nodata                    & 6057  &$   -67.0$&$  -20.2$ & 0.35&  5& \nodata                      & 2002-09-27  \\
\nodata                    & 6059  &$   -58.2$&$  -11.4$ & 0.57&  7& \nodata                      & 2002-09-27  \\

$0329379+034422$\tablenotemark{a} & \nodata      &$\mathbf{  -148.9}$&$ -160.9$ & 0.64&  7& 7.7           & 2002-09-28  \\
\nodata                    & 7075  &$  -154.3$&$ -166.3$ & 0.55&  7& \nodata                      & 2002-09-28  \\
\nodata                    & 7077  &$  -143.4$&$ -155.4$ & 0.73&  7& \nodata                      & 2002-09-28  \\

$0305165+200436$\tablenotemark{c} & \nodata      &$\mathbf{  -249.2}$&$ -196.5$ & 0.57&  7& 8.1                   & 2002-09-30  \\
\nodata                    & 9063  &$  -239.9$&$ -187.2$ & 0.49&  7& \nodata                      & 2002-09-30  \\
\nodata                    & 9065  &$  -253.9$&$ -201.2$ & 0.64&  7& \nodata                      & 2002-09-30  \\

$2309567-325813$\tablenotemark{b} & \nodata      &$\mathbf{   117.0}$&$  132.4$ & 0.36&  7&  11.3                 & 2002-09-30  \\
\nodata                    & 9048  &$   125.0$&$  140.4$ & 0.36&  5& \nodata                      & 2002-09-30  \\
\nodata                    & 9050  &$   109.0$&$  124.3$ & 0.35&  6& \nodata                      & 2002-09-30  \\
&&&&&&&\\

$0231171-190755$\tablenotemark{a} & \nodata      &$\mathbf{   -55.6}$&$  -96.6$ & 0.89&  7&   0.6                  & 2002-11-21/23  \\
\nodata                    & 3021  &$   -55.2$&$  -96.1$ & 0.59&  7& \nodata                   & 2002-11-21  \\
\nodata                    & 5009  &$   -56.0$&$  -97.0$ & 0.67&  7& \nodata                   & 2002-11-23  \\




\enddata
\tablenotetext{a}{All measures with CCP$\geq$0.5 have equal weight in average.}
\tablenotetext{b}{All measures with 0.3$<$CCP$<$0.5 have equal weight in average.}
\tablenotetext{c}{For those pairs having one measure with CCP$\geq$0.5 and one with 0.3$<$CCP$<$0.5, the former are given double weight compared to the latter.}
\tablenotetext{d}{For those pairs having one measure from either 2002-07-31 and 2002-08-01 (taken near the Ca IR triplet), and one measure taken on any other night (with the normal spectrograph setup) a straight average is taken, because the CCP scales differ between the two spectrograph setups.}
\end{deluxetable}

\oddsidemargin-0.5in
\begin{deluxetable}{rrrrrrrrrrrrrl}
\tabletypesize{\footnotesize}
\tablewidth{0pt}
\tablecaption{M Giants with Radial Velocity Data}
\tablehead{
\colhead{NAME}&     \colhead{$K_{s,o}$}&    \colhead{$\!\!\!(J$-$K_s)_o\!\!\!$}&   \colhead{$E_{B-V}$}&   \colhead{$l$}&      \colhead{$b$}&      \colhead{$d$}&  \colhead{$\Lambda_{\sun}$}& \colhead{$\!\!\!\!Z_{Sgr,GC}\!\!\!\!$}& \colhead{$v_{hel}$} &  \colhead{$v_{GSR}$}&  \colhead{$CCP$} & \colhead{$Q$} & \colhead{RV date} \\
\colhead{} & \colhead{}    & \colhead{}   & \colhead{}  &   \colhead{(deg)}&      \colhead{(deg)}&      \colhead{(kpc)}&  \colhead{(deg)}& \colhead{(kpc)}& \colhead{$\!\!\!$(km s$^{-1}$)$\!\!\!$} &  \colhead{$\!\!\!$(km s$^{-1}$)$\!\!\!$} &  \colhead{} & \colhead{} & \colhead{} }

\startdata

$0000571-295428$&  9.17&  1.07& 0.02&  15.9& $ -78.6$&   9.9&  64.5&  $  -1.0$& $   64.0$& $   71.5$& 0.63& 7& 2002-09-23 \\
$0002456-073004$&  8.81&  1.03& 0.03&  90.2& $ -67.2$&   7.0&  74.5&  $   2.0$& $  -52.3$& $   30.8$& 0.90& 7& 2002-11-26 \\
$0003163-442908$&  9.92&  1.02& 0.01& 328.9& $ -70.1$&  11.2&  58.1&  $  -3.7$& $    5.1$& $  -39.3$& 0.82& 7& 2002-11-26 \\
$0003528-194047$& 10.97&  1.03& 0.02&  64.8& $ -76.8$&  19.1&  69.5&  $   0.3$& $  -56.3$& $  -14.3$& 0.70& 7& 2002-11-26 \\
$0004321-711635$&  8.48&  0.96& 0.03& 308.3& $ -45.4$&   4.6&  40.4&  $  -2.4$& $   23.3$& $ -105.4$& 0.83& 7& 2002-09-29 \\
$0006594-275714$& 12.15&  0.97& 0.02&  25.6& $ -80.1$&  26.0&  66.6&  $  -3.5$& $  140.6$& $  152.4$& 0.70& 7& 2002-09-28 \\
$0007318-195427$&  8.60&  1.10& 0.02&  66.3& $ -77.6$&   8.4&  70.2&  $   0.5$& $ -187.3$& $ -147.8$& 0.76& 7& 2002-07-30 \\
$0010248+142130$&  8.29&  0.98& 0.06& 108.2& $ -47.3$&   4.5&  86.9&  $   3.0$& $  -65.0$& $   77.2$& 0.67& 7& 2002-09-26 \\
$0010323-025706$&  9.68&  1.08& 0.04&  99.2& $ -63.9$&  12.6&  78.3&  $   3.6$& $ -154.3$& $  -60.5$& 0.56& 7& 2002-09-23 \\
$0011085-712536$&  8.84&  0.96& 0.03& 307.5& $ -45.3$&   5.3&  40.8&  $  -3.0$& $   39.0$& $  -91.2$& 0.80& 7& 2002-09-29 \\
$0013189-230153$& 10.59&  1.00& 0.02&  56.2& $ -80.4$&  14.5&  70.1&  $  -0.6$& $   -6.5$& $   19.2$& 0.46& 7& 2002-07-31 \\
$0013562-172155$& 11.52&  0.99& 0.03&  79.4& $ -76.9$&  21.4&  72.7&  $   0.6$& $  -87.7$& $  -42.6$& 0.92& 7& 2002-07-19 \\
$0016009-045246$&  8.93&  1.01& 0.03& 100.6& $ -66.2$&   6.9&  78.7&  $   2.1$& $ -189.3$& $ -104.4$& 1.04& 7& 2002-11-26 \\
$0017021+010417$& 10.89&  0.95& 0.03& 105.2& $ -60.6$&  13.6&  81.7&  $   4.5$& $ -219.2$& $ -116.6$& 0.73& 7& 2002-09-23 \\
$0022356-051208$&  9.81&  1.08& 0.03& 104.3& $ -67.0$&  13.9&  80.0&  $   3.1$& $ -121.7$& $  -41.2$& 0.55& 7& 2002-09-23 \\
$0022578-284534$&  9.06&  0.95& 0.01&  16.7& $ -83.5$&   5.8&  69.4&  $  -0.3$& $  -31.4$& $  -29.8$& 0.80& 7& 2002-09-29 \\
$0023424-553344$&  8.84&  1.03& 0.01& 311.1& $ -61.1$&   7.2&  54.9&  $  -3.3$& $   85.2$& $   -2.3$& 0.75& 7& 2002-09-24 \\
$0026467-152643$& 10.82&  0.95& 0.03&  95.6& $ -77.0$&  13.2&  76.3&  $   0.8$& $  -64.3$& $  -19.3$& 0.60& 7& 2002-09-23 \\
$0029133-375036$&  9.78&  1.05& 0.02& 325.1& $ -78.3$&  12.1&  66.2&  $  -3.3$& $  -72.5$& $ -104.7$& 0.84& 7& 2002-11-21 \\
$0030183-463120$&  9.27&  1.01& 0.01& 313.7& $ -70.2$&   7.9&  61.6&  $  -2.8$& $  -54.4$& $ -115.7$& 0.87& 7& 2002-11-24 \\
$0032165-185111$& 11.57&  1.02& 0.02&  93.9& $ -80.6$&  24.5&  75.9&  $  -0.9$& $  -73.8$& $  -43.1$& 0.68& 7& 2002-09-27 \\
$0034388+021838$&  9.04&  0.98& 0.02& 114.4& $ -60.3$&   6.4&  86.3&  $   2.5$& $  -13.8$& $   82.8$& 0.76& 7& 2002-09-26 \\
$0034522-213714$&  8.48&  0.99& 0.02&  87.6& $ -83.3$&   5.2&  75.2&  $   0.3$& $   40.2$& $   60.2$& 0.86& 7& 2002-09-29 \\
$0035110-584639$&  8.33&  0.99& 0.01& 306.9& $ -58.2$&   4.8&  54.2&  $  -2.1$& $  114.3$& $   13.6$& 0.84& 7& 2002-09-24 \\
$0039096-132243$& 10.50&  1.12& 0.02& 110.5& $ -76.0$&  21.9&  79.9&  $   0.9$& $  -95.8$& $  -50.6$& 1.19& 7& 2002-07-31 \\
$0039567-290607$&  9.17&  0.95& 0.02& 354.3& $ -86.8$&   6.1&  72.6&  $  -0.6$& $  -21.3$& $  -49.6$& 0.73& 7& 2002-09-29 \\
$0040562-222255$&  8.77&  1.01& 0.02&  95.7& $ -84.7$&   6.5&  76.1&  $  -0.0$& $  -56.6$& $  -42.3$& 0.92& 7& 2002-07-30 \\
$0041193-192518$&  8.84&  0.95& 0.02& 105.6& $ -82.0$&   5.2&  77.5&  $   0.4$& $   67.3$& $   91.4$& 0.73& 7& 2002-09-26 \\
$0042438-222816$& 11.06&  1.02& 0.02&  99.5& $ -84.9$&  19.3&  76.4&  $  -1.9$& $ -204.0$& $ -119.1$& 1.25& 7& 2002-11-26 \\
$0042542-562702$&  8.19&  0.97& 0.02& 305.3& $ -60.6$&   4.1&  57.0&  $  -1.6$& $  181.7$& $   85.4$& 1.04& 7& 2002-09-24 \\
$0046441-065926$& 11.84&  0.95& 0.06& 119.5& $ -69.8$&  21.0&  84.5&  $   2.7$& $ -124.8$& $  -63.3$& 0.78& 7& 2002-09-27 \\
$0048046-113155$& 10.61&  1.05& 0.03& 119.9& $ -74.4$&  17.3&  82.7&  $   1.1$& $  -88.3$& $  -42.1$& 0.94& 7& 2002-07-16 \\
$0048087-135541$&  9.67&  1.06& 0.02& 119.4& $ -76.8$&  11.9&  81.6&  $   0.6$& $   -7.4$& $   30.8$& 0.65& 7& 2002-11-25 \\
$0049581-334807$&  8.66&  1.02& 0.01& 305.6& $ -83.3$&   6.3&  72.2&  $  -1.2$& $  -31.4$& $  -59.7$& 0.79& 7& 2002-09-29 \\
$0052180-545157$&  8.65&  0.99& 0.02& 302.7& $ -62.3$&   5.6&  59.5&  $  -2.5$& $  107.0$& $   12.1$& 0.89& 7& 2002-09-24 \\
$0052298-151836$& 11.16&  1.08& 0.02& 124.2& $ -78.2$&  25.3&  81.9&  $  -0.5$& $  -86.7$& $  -55.3$& 0.70& 7& 2002-11-25 \\
$0053201-052948$& 10.46&  1.01& 0.04& 124.2& $ -68.4$&  14.0&  86.7&  $   2.2$& $ -152.1$& $  -89.8$& 1.18& 7& 2002-07-31 \\
$0053545-675121$&  8.72&  0.97& 0.02& 302.6& $ -49.3$&   5.3&  48.4&  $  -3.0$& $  120.8$& $   -8.6$& 0.96& 7& 2002-09-29 \\
$0054207-044917$& 10.51&  1.01& 0.05& 124.8& $ -67.7$&  14.3&  87.3&  $   2.3$& $ -139.9$& $  -76.0$& 0.68& 7& 2002-09-27 \\
$0055445-403004$&  8.95&  1.00& 0.01& 299.4& $ -76.6$&   6.7&  69.5&  $  -2.0$& $  107.1$& $   54.5$& 0.89& 7& 2002-09-29 \\
$0056333-215439$& 10.74&  1.06& 0.02& 135.8& $ -84.7$&  19.2&  79.5&  $  -2.3$& $  -48.3$& $  -40.8$& 0.77& 7& 2002-11-25 \\
$0057247-001845$&  9.70&  0.97& 0.03& 126.2& $ -63.1$&   8.2&  90.1&  $   2.2$& $   32.8$& $  108.7$& 0.61& 7& 2002-09-23 \\
$0058503+071403$&  9.08&  0.95& 0.07& 126.2& $ -55.6$&   5.9&  94.2&  $   2.5$& $  -80.8$& $   16.0$& 0.86& 7& 2002-09-26 \\
$0058567-373343$&  9.63&  1.00& 0.01& 294.8& $ -79.4$&   9.0&  71.8&  $  -2.7$& $  -86.6$& $ -131.4$& 0.79& 7& 2002-11-26 \\
$0101193-153634$& 11.20&  1.01& 0.02& 134.7& $ -78.3$&  19.6&  83.6&  $  -0.6$& $  -82.2$& $  -56.8$& 0.85& 7& 2002-09-27 \\
$0101403-300004$& 10.57&  0.95& 0.02& 265.5& $ -86.4$&  11.7&  76.4&  $  -2.5$& $  -35.4$& $  -57.2$& 0.60& 7& 2002-09-19 \\
$0101538-101508$& 11.31&  1.01& 0.03& 131.7& $ -72.9$&  21.0&  86.3&  $   0.9$& $ -130.9$& $  -88.6$& 0.73& 7& 2002-11-26 \\
$0104102-142958$& 10.87&  1.07& 0.02& 136.8& $ -77.0$&  21.6&  84.7&  $  -0.6$& $  -86.1$& $  -58.8$& 0.94& 7& 2002-07-31 \\
$0104400-664135$&  8.13&  0.95& 0.02& 300.9& $ -50.4$&   3.8&  50.7&  $  -1.9$& $  106.1$& $  -23.0$& 1.01& 7& 2002-09-29 \\
$0104589-405126$&  8.35&  1.03& 0.01& 292.3& $ -76.0$&   5.7&  70.9&  $  -1.7$& $   97.6$& $   39.7$& 0.91& 7& 2002-09-29 \\
$0105093+143325$&  8.79&  1.01& 0.05& 127.9& $ -48.2$&   6.4&  99.6&  $   3.2$& $ -141.9$& $  -28.8$& 0.73& 7& 2002-09-26 \\
$0108354-412937$&  8.81&  0.99& 0.01& 290.3& $ -75.2$&   5.9&  71.2&  $  -1.8$& $    3.8$& $  -57.5$& 0.74& 7& 2002-09-29 \\
$0109191-150816$&  9.27&  1.14& 0.03& 143.0& $ -77.3$&  13.6&  85.5&  $  -0.3$& $  -82.8$& $  -60.5$& 0.21& 5& 2002-07-30 \\
$0111418-110857$&  9.38&  0.95& 0.02& 140.5& $ -73.3$&   6.8&  88.0&  $   0.7$& $  162.5$& $  196.1$& 0.98& 7& 2002-09-20 \\
$0111570-043342$& 11.17&  1.00& 0.05& 136.1& $ -66.9$&  18.6&  91.2&  $   2.2$& $ -132.1$& $  -78.0$& 0.80& 7& 2002-07-16 \\
$0114396-084315$& 11.21&  1.02& 0.05& 140.6& $ -70.8$&  20.4&  89.8&  $   0.9$& $ -114.7$& $  -75.3$& 0.45& 7& 2002-09-20 \\
$0115044-592237$&  8.27&  1.05& 0.02& 297.3& $ -57.5$&   5.9&  58.9&  $  -3.1$& $   78.0$& $  -36.2$& 0.79& 7& 2002-09-24 \\
$0118282-304045$& 10.01&  1.04& 0.02& 245.5& $ -83.1$&  12.9&  79.3&  $  -3.4$& $  -80.5$& $ -113.3$& 0.70& 7& 2002-09-19 \\
$0119125-523316$&  8.26&  1.00& 0.01& 293.3& $ -64.0$&   4.9&  65.3&  $  -2.1$& $   92.2$& $   -5.6$& 0.89& 7& 2002-09-24 \\
$0119194+003404$& 11.47&  0.98& 0.03& 137.7& $ -61.5$&  19.3&  95.4&  $   3.4$& $ -159.4$& $  -94.3$& 0.85& 7& 2002-09-27 \\
$0120290-024340$& 10.92&  1.07& 0.05& 140.1& $ -64.6$&  21.9&  94.0&  $   2.6$& $ -158.2$& $ -103.7$& 0.69& 7& 2002-09-20 \\
$0121232-103610$& 10.06&  1.08& 0.03& 147.5& $ -72.0$&  15.5&  90.3&  $   0.2$& $ -126.2$& $  -96.7$& 0.91& 7& 2002-07-16 \\
$0121255-154349$& 10.42&  1.14& 0.02& 155.9& $ -76.6$&  22.6&  87.8&  $  -1.9$& $  -75.3$& $  -62.1$& 0.92& 7& 2002-07-31 \\
$0123350-053817$& 11.54&  1.01& 0.03& 144.0& $ -67.2$&  23.0&  93.2&  $   1.5$& $ -159.1$& $ -115.5$& 0.75& 7& 2002-09-27 \\
$0123515-000144$&  8.39&  0.95& 0.03& 140.3& $ -61.8$&   4.3&  96.1&  $   1.4$& $ -187.0$& $ -126.4$& 1.19& 7& 2002-09-26 \\
$0125348-512246$&  8.87&  1.00& 0.01& 290.3& $ -64.9$&   6.4&  67.1&  $  -3.0$& $   81.5$& $  -15.6$& 0.77& 7& 2002-09-24 \\
$0126160-240307$&  9.43&  1.01& 0.01& 193.5& $ -81.6$&   8.7&  84.5&  $  -1.3$& $  115.0$& $   98.8$& 1.04& 7& 2002-11-26 \\
$0128181-303622$&  9.39&  1.01& 0.02& 238.4& $ -81.2$&   8.5&  81.3&  $  -2.1$& $   -1.4$& $  -39.2$& 0.74& 7& 2002-09-19 \\
$0128276-050517$& 11.33&  0.96& 0.04& 146.4& $ -66.3$&  17.1&  94.6&  $   1.3$& $ -121.7$& $  -79.5$& 0.80& 7& 2002-09-28 \\
$0129496-045107$& 10.87&  1.02& 0.03& 147.0& $ -65.9$&  17.5&  95.0&  $   1.3$& $ -111.4$& $  -69.2$& 0.80& 7& 2002-07-19 \\
$0130267-253541$&  8.85&  1.00& 0.01& 205.2& $ -81.1$&   6.3&  84.5&  $  -0.9$& $  238.9$& $  215.5$& 1.08& 7& 2002-09-29 \\
$0130289-693029$&  8.84&  0.96& 0.02& 297.9& $ -47.2$&   5.5&  49.9&  $  -3.4$& $  150.2$& $    8.8$& 0.46& 7& 2002-09-24 \\
$0133268-295719$&  9.39&  1.04& 0.02& 232.4& $ -80.4$&   9.7&  82.7&  $  -2.5$& $  123.7$& $   85.1$& 0.88& 7& 2002-11-23 \\
$0138353-030718$& 10.95&  1.07& 0.03& 150.1& $ -63.5$&  22.0&  97.7&  $   1.6$& $ -124.5$& $  -82.6$& 0.34& 7& 2002-09-23 \\
$0138357-143354$&  8.80&  1.09& 0.02& 166.5& $ -73.3$&   8.8&  92.0&  $  -0.3$& $   27.6$& $   34.0$& 0.71& 7& 2002-09-19 \\
$0140241-201259$& 10.87&  1.02& 0.02& 183.9& $ -76.8$&  17.4&  89.4&  $  -3.1$& $  207.9$& $  195.5$& 0.73& 7& 2002-11-24 \\
$0142411-111317$& 10.69&  1.07& 0.02& 162.5& $ -70.0$&  19.7&  94.5&  $  -1.0$& $ -126.0$& $ -111.8$& 0.53& 7& 2002-09-23 \\
$0146287-032119$& 11.60&  1.02& 0.02& 154.3& $ -62.8$&  24.1&  99.3&  $   1.2$& $ -127.7$& $  -91.7$& 0.79& 7& 2002-09-28 \\
$0148419-010653$& 10.66&  1.01& 0.03& 153.2& $ -60.6$&  15.5& 100.9&  $   1.5$& $ -165.5$& $ -124.1$& 0.86& 7& 2002-07-31 \\
$0151211-072745$& 11.69&  1.01& 0.02& 161.5& $ -65.7$&  24.4&  98.3&  $  -0.6$& $ -121.2$& $ -100.9$& 0.77& 7& 2002-09-28 \\
$0157416-170947$& 10.65&  1.05& 0.02& 183.3& $ -71.7$&  17.5&  94.6&  $  -2.9$& $  -86.5$& $ -100.2$& 0.60& 7& 2002-09-23 \\
$0159361-080113$& 10.66&  1.05& 0.02& 166.3& $ -65.0$&  18.0&  99.8&  $  -0.7$& $ -112.5$& $  -99.3$& 1.08& 7& 2002-07-31 \\
$0202043-074527$& 11.16&  1.01& 0.02& 167.0& $ -64.4$&  19.6& 100.4&  $  -0.9$& $ -157.2$& $ -144.9$& 0.74& 7& 2002-09-20 \\
$0202225+003225$&  8.59&  1.04& 0.02& 157.5& $ -57.5$&   6.7& 104.7&  $   1.1$& $  133.1$& $  170.5$& 0.46& 7& 2002-09-26 \\
$0203463+080420$& 11.36&  1.02& 0.08& 151.9& $ -50.7$&  22.2& 108.9&  $   4.1$& $ -187.9$& $ -129.3$& 0.69& 7& 2002-09-30 \\
$0206248+092837$& 11.87&  0.96& 0.07& 151.9& $ -49.1$&  22.0& 110.2&  $   4.4$& $ -190.5$& $ -129.6$& 0.76& 7& 2002-09-30 \\
$0208007+054857$&  8.70&  0.96& 0.06& 155.1& $ -52.2$&   5.1& 108.6&  $   1.4$& $  114.0$& $  163.3$& 0.86& 7& 2002-09-26 \\
$0210191+083551$& 11.28&  0.92& 0.06& 153.8& $ -49.5$&  14.3& 110.5&  $   2.9$& $ -184.6$& $ -128.8$& 0.63& 7& 2002-07-19 \\
$0212569-052952$& 10.64&  1.03& 0.03& 168.5& $ -60.9$&  16.3& 103.9&  $  -0.4$& $ -125.7$& $ -113.6$& 1.21& 7& 2002-07-31 \\
$0213311-111530$&  8.86&  1.08& 0.02& 177.4& $ -65.0$&   8.7& 101.1&  $  -0.6$& $  127.1$& $  121.3$& 0.51& 7& 2002-07-30 \\
$0215595-003221$& 11.37&  0.96& 0.04& 163.8& $ -56.5$&  17.1& 107.1&  $   0.7$& $ -171.4$& $ -146.4$& 0.94& 7& 2002-09-30 \\
$0218528-490421$&  8.37&  1.01& 0.02& 271.2& $ -62.3$&   5.2&  77.4&  $  -2.5$& $  115.4$& $    1.6$& 0.88& 7& 2002-09-24 \\
$0219409+110359$& 10.23&  1.01& 0.08& 155.1& $ -46.2$&  12.6& 113.8&  $   2.8$& $   12.6$& $   69.3$& 0.62& 7& 2002-11-24 \\
$0220578-203941$&  9.59&  0.97& 0.02& 200.2& $ -68.6$&   7.9&  97.6&  $  -1.6$& $   47.4$& $    8.5$& 0.81& 7& 2002-09-19 \\
$0221321-120741$&  9.10&  0.98& 0.02& 182.0& $ -64.1$&   6.7& 102.3&  $  -0.4$& $    3.2$& $  -10.3$& 0.72& 7& 2002-09-23 \\
$0222230+013319$& 11.59&  1.04& 0.04& 163.9& $ -53.9$&  26.5& 109.5&  $   1.1$& $ -142.4$& $ -115.4$& 0.89& 7& 2002-09-28 \\
$0222294-204354$&  8.85&  0.98& 0.03& 200.7& $ -68.3$&   5.9&  97.8&  $  -1.0$& $   52.0$& $   11.9$& 0.79& 7& 2002-09-29 \\
$0223373-004200$& 10.95&  1.03& 0.04& 166.7& $ -55.5$&  19.3& 108.7&  $   0.3$& $ -181.0$& $ -161.6$& 0.61& 7& 2002-09-20 \\
$0226332-005318$& 10.60&  1.00& 0.03& 167.9& $ -55.2$&  14.3& 109.2&  $   0.3$& $ -120.6$& $ -103.8$& 0.70& 7& 2002-11-25 \\
$0228464-002133$& 11.48&  1.00& 0.03& 168.1& $ -54.5$&  21.1& 109.9&  $   0.1$& $ -150.1$& $ -133.2$& 0.96& 7& 2002-09-28 \\
$0230079+050840$&  9.09&  1.00& 0.07& 163.1& $ -49.9$&   7.0& 113.0&  $   1.2$& $  -36.0$& $   -3.5$& 0.79& 7& 2002-09-26 \\
$0230418-520141$&  9.03&  0.96& 0.03& 272.9& $ -58.9$&   6.0&  76.5&  $  -3.3$& $  388.5$& $  263.2$& 0.84& 7& 2002-09-24 \\
$0231171-190755$&  9.34&  1.03& 0.03& 199.0& $ -65.7$&   9.2& 100.6&  $  -2.0$& $  -55.6$& $  -96.6$& 0.89& 7& 2002-11-21/23 \\
$0232358-255822$&  8.87&  0.98& 0.02& 215.8& $ -67.4$&   5.9&  96.9&  $  -1.5$& $  -62.0$& $ -123.5$& 0.80& 7& 2002-09-29 \\
$0233240+102510$& 10.71&  1.12& 0.11& 159.7& $ -45.0$&  24.0& 116.4&  $   3.7$& $ -177.5$& $ -131.8$& 0.97& 7& 2002-07-31 \\
$0233246-375454$&  8.14&  1.02& 0.02& 246.1& $ -66.1$&   4.9&  89.0&  $  -1.9$& $  215.0$& $  121.0$& 1.03& 7& 2002-09-29 \\
$0235206-094228$&  8.54&  0.99& 0.03& 182.5& $ -59.9$&   5.3& 106.6&  $  -0.1$& $  -52.2$& $  -67.8$& 0.82& 7& 2002-09-26 \\
$0237111-175444$& 10.19&  0.97& 0.02& 197.8& $ -64.0$&  10.3& 102.5&  $  -2.3$& $  123.1$& $   82.0$& 1.05& 7& 2002-11-26 \\
$0237542-100623$& 10.40&  1.00& 0.03& 183.9& $ -59.7$&  12.8& 106.9&  $  -1.7$& $ -170.0$& $ -188.5$& 0.54& 7& 2002-09-19 \\
$0239004-060541$& 11.04&  1.00& 0.03& 178.4& $ -56.9$&  17.7& 109.2&  $  -1.7$& $ -124.9$& $ -132.1$& 0.79& 7& 2002-09-23 \\
$0239543-650056$&  8.43&  0.98& 0.03& 286.1& $ -48.4$&   4.8&  62.1&  $  -2.9$& $   79.2$& $  -72.1$& 0.87& 7& 2002-09-24 \\
$0239597-265452$&  9.65&  1.01& 0.02& 218.7& $ -65.9$&   9.8&  97.8&  $  -3.4$& $  100.4$& $   31.8$& 0.88& 7& 2002-11-26 \\
$0240272+063023$&  8.30&  0.99& 0.10& 165.0& $ -47.3$&   4.7& 115.9&  $   1.1$& $  -62.6$& $  -33.1$& 0.90& 7& 2002-09-26 \\
$0241453-300637$&  8.64&  0.95& 0.02& 226.5& $ -65.7$&   4.8&  96.1&  $  -1.4$& $  107.0$& $   28.8$& 0.87& 7& 2002-09-23 \\
$0243431+123247$& 12.46&  1.00& 0.11& 161.0& $ -41.9$&  33.6& 119.7&  $   5.1$& $ -193.2$& $ -148.1$& 0.63& 7& 2002-09-30 \\
$0247498+011136$& 11.42&  1.06& 0.05& 172.3& $ -50.2$&  26.4& 114.8&  $  -0.6$& $ -139.5$& $ -130.8$& 0.46& 7& 2002-09-27 \\
$0248166-493405$&  8.98&  0.98& 0.03& 265.9& $ -58.3$&   6.1&  81.7&  $  -3.4$& $   43.1$& $  -84.5$& 0.72& 7& 2002-09-24 \\
$0248456-243409$&  9.73&  1.01& 0.01& 214.0& $ -63.5$&   9.9& 101.1&  $  -3.3$& $  102.4$& $   35.0$& 1.04& 7& 2002-11-21 \\
$0251579-645410$&  8.36&  1.01& 0.03& 284.5& $ -47.7$&   5.4&  63.4&  $  -3.5$& $  153.4$& $   -1.3$& 0.90& 7& 2002-09-24 \\
$0255431+033923$& 11.57&  0.99& 0.09& 172.0& $ -47.0$&  21.4& 117.8&  $   0.1$& $ -161.6$& $ -151.0$& 0.61& 7& 2002-09-27 \\
$0256141-591714$&  8.63&  0.99& 0.02& 277.9& $ -51.3$&   5.6&  71.7&  $  -3.5$& $  125.3$& $  -22.7$& 0.84& 7& 2002-09-24 \\
$0258253-002645$&  8.76&  0.98& 0.11& 177.0& $ -49.5$&   5.6& 116.3&  $   0.3$& $   -6.7$& $  -10.2$& 0.79& 7& 2002-09-26 \\
$0300064+133705$&  9.49&  1.03& 0.15& 164.4& $ -38.7$&   9.9& 123.7&  $   1.9$& $    0.0$& $   37.1$& 0.81& 7& 2002-11-24 \\
$0300127+055044$& 10.68&  1.05& 0.10& 171.1& $ -44.7$&  18.1& 119.8&  $   0.7$& $ -149.2$& $ -135.0$& 1.08& 7& 2002-07-31 \\
$0304531-384409$&  8.73&  1.04& 0.02& 244.1& $ -59.9$&   7.0&  94.4&  $  -3.5$& $  -66.5$& $ -179.2$& 0.80& 7& 2002-09-29 \\
$0305165+200436$& 10.00&  1.03& 0.34& 160.9& $ -32.7$&  12.4& 128.1&  $   3.3$& $ -249.2$& $ -196.5$& 0.57& 7& 2002-09-30 \\
$0308171-042052$&  8.81&  0.99& 0.05& 184.1& $ -50.2$&   6.1& 116.5&  $  -0.2$& $  -72.3$& $  -94.1$& 0.81& 7& 2002-09-26 \\
$0310455+184214$& 11.06&  0.99& 0.10& 163.1& $ -33.1$&  17.1& 128.5&  $   3.6$& $ -215.0$& $ -169.8$& 0.92& 7& 2002-09-30 \\
$0313549-135514$& 10.35&  1.00& 0.04& 198.5& $ -54.3$&  12.7& 112.7&  $  -3.3$& $   35.9$& $  -17.6$& 0.72& 7& 2002-11-21 \\
$0314413-200034$&  8.97&  0.97& 0.03& 208.5& $ -56.5$&   6.0& 109.4&  $  -1.6$& $   70.8$& $   -0.2$& 0.73& 7& 2002-09-29 \\
$0315440-372611$&  8.84&  0.98& 0.03& 240.9& $ -58.0$&   5.9&  97.6&  $  -2.8$& $   89.0$& $  -26.4$& 0.64& 7& 2002-09-29 \\
$0316391-070637$& 10.10&  0.97& 0.07& 189.5& $ -50.2$&  10.1& 116.9&  $  -1.5$& $   29.0$& $   -6.5$& 0.83& 7& 2002-11-25 \\
$0317036+030631$& 11.63&  1.02& 0.11& 177.9& $ -43.7$&  25.0& 122.2&  $  -1.3$& $ -113.0$& $ -118.5$& 0.47& 7& 2002-09-27 \\
$0317573+030345$&  9.98&  1.00& 0.14& 178.2& $ -43.5$&  10.6& 122.3&  $  -0.1$& $    2.3$& $   -3.7$& 0.87& 7& 2002-11-26 \\
$0318182-003128$&  8.54&  0.99& 0.06& 182.1& $ -45.9$&   5.3& 120.6&  $   0.1$& $   61.6$& $   44.5$& 0.76& 7& 2002-09-26 \\
$0318549-255739$&  8.75&  0.95& 0.01& 219.4& $ -57.1$&   5.1& 106.6&  $  -1.7$& $   29.1$& $  -60.5$& 0.72& 7& 2002-09-29 \\
$0322152+164323$& 11.03&  1.00& 0.16& 167.2& $ -32.9$&  17.6& 129.9&  $   2.8$& $  -27.1$& $    4.7$& 0.64& 7& 2002-09-28 \\
$0326123+175628$& 12.50&  0.97& 0.11& 167.1& $ -31.4$&  29.9& 131.3&  $   4.4$& $ -186.3$& $ -153.3$& 0.58& 7& 2002-09-30 \\
$0327007-012321$& 11.79&  0.97& 0.08& 185.0& $ -44.7$&  21.5& 122.1&  $  -2.9$& $ -172.5$& $ -198.2$& 0.36& 4& 2002-09-20 \\
$0329379+034422$& 12.26&  0.96& 0.14& 180.2& $ -40.9$&  26.3& 125.2&  $  -1.9$& $ -148.9$& $ -159.0$& 0.67& 7& 2002-09-28 \\
$0329408-302652$&  9.61&  0.95& 0.01& 228.0& $ -55.3$&   7.5& 105.8&  $  -3.4$& $   36.0$& $  -71.1$& 0.75& 7& 2002-09-26 \\
$0331233-144614$&  8.27&  1.02& 0.06& 202.8& $ -50.9$&   5.2& 116.1&  $  -1.1$& $  -36.0$& $ -103.5$& 0.79& 7& 2002-09-26 \\
$0331242-053102$&  8.46&  0.97& 0.06& 190.6& $ -46.3$&   4.8& 121.0&  $  -0.3$& $ -243.3$& $ -284.2$& 0.95& 7& 2002-09-26 \\
$0332064-474215$&  8.43&  0.96& 0.01& 257.2& $ -52.7$&   4.5&  91.0&  $  -2.5$& $  169.2$& $   25.2$& 0.95& 7& 2002-09-24 \\
$0332342-252639$&  8.65&  0.98& 0.02& 219.6& $ -54.0$&   5.4& 109.9&  $  -1.9$& $  -66.2$& $ -162.8$& 0.69& 7& 2002-09-26 \\
$0333157-012641$& 11.02&  1.06& 0.10& 186.4& $ -43.5$&  22.1& 123.5&  $  -3.3$& $ -139.4$& $ -169.5$& 0.76& 7& 2002-11-24 \\
$0333199-341700$&  8.85&  0.98& 0.01& 234.8& $ -54.7$&   5.8& 103.7&  $  -2.8$& $  154.6$& $   36.4$& 1.00& 7& 2002-09-26 \\
$0334210+051809$& 11.28&  1.02& 0.26& 179.7& $ -39.0$&  21.2& 127.0&  $  -1.1$& $ -154.6$& $ -165.0$& 0.35& 7& 2002-09-27 \\
$0340164+090338$& 11.66&  1.03& 0.35& 177.4& $ -35.4$&  26.1& 130.1&  $  -0.3$& $ -140.3$& $ -143.2$& 0.39& 6& 2002-11-26 \\
$0342036-540712$&  8.59&  0.97& 0.02& 265.8& $ -48.9$&   5.1&  84.8&  $  -3.2$& $  156.4$& $   -1.3$& 1.01& 7& 2002-09-24 \\
$0342225+054745$& 10.77&  1.06& 0.22& 180.9& $ -37.2$&  19.6& 129.0&  $  -1.1$& $ -156.1$& $ -170.4$& 0.55& 7& 2002-09-28 \\
$0342465+074646$&  9.51&  1.07& 0.28& 179.1& $ -35.8$&  11.4& 130.0&  $   0.1$& $   87.0$& $   78.4$& 0.65& 7& 2002-11-23 \\
$0348353-122557$&  9.82&  0.97& 0.05& 202.4& $ -46.1$&   8.8& 121.3&  $  -2.5$& $   38.2$& $  -33.7$& 0.80& 7& 2002-09-26 \\
$0348437+065236$& 10.79&  1.02& 0.20& 181.1& $ -35.3$&  17.1& 130.9&  $  -0.8$& $ -154.7$& $ -169.9$& 0.61& 7& 2002-11-25 \\
$0349517-422532$&  8.19&  1.05& 0.01& 247.8& $ -50.8$&   5.8&  99.9&  $  -3.4$& $  160.5$& $   17.3$& 0.79& 7& 2002-09-24 \\
$0349531+093613$& 12.11&  0.97& 0.17& 178.8& $ -33.3$&  25.5& 132.4&  $  -0.6$& $ -128.4$& $ -136.0$& 0.64& 7& 2002-09-30 \\
$0350029-310009$&  8.43&  1.04& 0.01& 229.6& $ -51.0$&   6.2& 109.8&  $  -3.0$& $  307.2$& $  186.9$& 1.09& 7& 2002-09-26 \\
$0357262+053248$& 10.07&  1.09& 0.29& 184.0& $ -34.5$&  15.9& 132.2&  $  -1.2$& $ -144.2$& $ -169.2$& 0.53& 7& 2002-11-23 \\
$0358367+071547$& 11.87&  1.04& 0.20& 182.7& $ -33.1$&  30.1& 133.3&  $  -2.5$& $   26.0$& $    5.5$& 0.83& 7& 2002-11-25 \\
$0359545-162256$&  8.83&  0.96& 0.04& 209.1& $ -45.2$&   5.4& 121.6&  $  -1.6$& $  139.8$& $   49.6$& 1.00& 7& 2002-09-26 \\
$0403354+055238$& 11.37&  1.02& 0.21& 184.9& $ -33.0$&  22.1& 133.8&  $  -2.3$& $   -7.1$& $  -35.0$& 0.45& 7& 2002-11-24 \\
$0404282-401854$&  9.00&  0.96& 0.01& 244.1& $ -48.3$&   5.9& 104.9&  $  -3.4$& $   70.1$& $  -76.4$& 0.71& 7& 2002-09-24 \\
$0404400-380337$&  8.33&  1.02& 0.01& 240.7& $ -48.3$&   5.5& 107.1&  $  -3.1$& $   20.2$& $ -122.3$& 0.80& 7& 2002-11-22 \\
$0405475-553719$&  8.06&  0.98& 0.01& 266.0& $ -45.1$&   4.2&  86.1&  $  -2.6$& $  152.3$& $  -16.1$& 1.04& 7& 2002-09-24 \\
$0408285+045043$& 11.34&  1.01& 0.40& 186.7& $ -32.7$&  21.1& 134.4&  $  -2.6$& $ -124.1$& $ -158.4$& 0.53& 7& 2002-11-23 \\
$0411368+140152$& 11.19&  1.02& 0.50& 179.0& $ -26.4$&  20.2& 139.2&  $   0.3$& $ -154.8$& $ -162.5$& 0.63& 7& 2002-11-26 \\
$0416369+081048$& 10.75&  1.00& 0.45& 185.0& $ -29.1$&  15.4& 137.7&  $  -1.1$& $ -112.9$& $ -142.1$& 0.31& 5& 2002-11-23 \\
$0416573+064649$& 11.07&  1.00& 0.26& 186.4& $ -29.9$&  18.0& 137.2&  $  -1.9$& $  -94.4$& $ -128.1$& 0.37& 7& 2002-11-25 \\
$0420006+224339$& 11.05&  1.05& 0.25& 173.4& $ -19.1$&  21.1& 144.8&  $   2.8$& $ -154.6$& $ -140.1$& 0.25& 6& 2002-11-25 \\
$0420154+080859$& 11.90&  1.00& 0.22& 185.7& $ -28.4$&  26.0& 138.5&  $  -2.7$& $ -128.3$& $ -159.9$& 0.61& 7& 2002-11-26 \\
$0422084+083418$& 11.33&  1.05& 0.24& 185.6& $ -27.8$&  24.7& 139.1&  $  -2.5$& $   22.7$& $   -8.6$& 0.57& 7& 2002-11-23 \\
$1307425+011347$&  8.78&  1.03& 0.03& 312.2& $  63.8$&   6.9& 271.9&  $   0.0$& $   41.2$& $  -25.5$& 0.69& 7& 2002-07-30 \\
$1309134+121550$& 10.36&  1.01& 0.02& 319.5& $  74.6$&  13.5& 266.9&  $   1.5$& $  -33.6$& $  -65.1$& 0.34& 4& 2002-07-31 \\
$1318500+061112$&  9.23&  1.02& 0.04& 321.4& $  68.1$&   8.3& 271.9&  $   0.7$& $   20.0$& $  -24.7$& 0.79& 7& 2002-07-30 \\
$1334591+043041$&  8.57&  1.01& 0.03& 329.5& $  65.1$&   5.7& 276.3&  $   0.8$& $  132.1$& $   92.2$& 1.01& 7& 2002-07-30 \\
$1346332+001009$&  9.72&  1.02& 0.03& 331.3& $  60.0$&  10.5& 280.9&  $   0.3$& $ -156.6$& $ -202.2$& 0.76& 7& 2002-07-30 \\
$1409365-013356$&  9.41&  1.01& 0.06& 339.3& $  55.7$&   8.4& 286.8&  $   0.6$& $   60.2$& $   24.6$& 0.73& 7& 2002-07-30 \\
$1411221-061013$&  9.51&  1.08& 0.04& 336.0& $  51.5$&  11.9& 289.5&  $  -0.3$& $   31.0$& $  -16.8$& 0.64& 7& 2002-07-30 \\
$1412322-052131$& 10.23&  1.03& 0.03& 337.1& $  52.1$&  13.4& 289.4&  $  -0.2$& $  -84.2$& $ -129.0$& 0.77& 7& 2002-07-31 \\
$1415453-100002$&  9.07&  1.06& 0.05& 334.6& $  47.6$&   8.8& 292.4&  $  -0.4$& $   89.1$& $   32.8$& 0.72& 7& 2002-07-30 \\
$1428255-082436$&  8.62&  1.04& 0.05& 339.8& $  47.4$&   6.8& 294.3&  $   0.2$& $    2.6$& $  -40.5$& 0.63& 7& 2002-07-31 \\
$1429162-075540$&  9.75&  1.02& 0.05& 340.5& $  47.7$&  10.5& 294.3&  $  -0.0$& $  -30.9$& $  -72.1$& 0.69& 7& 2002-07-30/31 \\
$1434332-014057$&  8.88&  1.03& 0.04& 347.8& $  52.0$&   7.3& 292.2&  $   1.0$& $   11.7$& $   -7.2$& 0.79& 7& 2002-07-30 \\
$1437449-093505$&  9.01&  1.07& 0.09& 341.7& $  45.1$&   9.1& 296.9&  $   0.1$& $  225.0$& $  184.6$& 0.87& 7& 2002-07-31 \\
$1440459+002945$& 11.86&  1.06& 0.05& 352.1& $  52.6$&  33.0& 292.5&  $   3.0$& $   15.8$& $    7.4$& 0.25& 4& 2002-08-01 \\
$1441233-124407$&  9.51&  1.06& 0.07& 340.2& $  42.0$&  11.0& 299.3&  $  -0.6$& $ -133.9$& $ -181.1$& 0.57& 7& 2002-07-30 \\
$1448055-034313$& 11.67&  1.14& 0.10& 349.8& $  48.2$&  40.6& 296.2&  $   1.6$& $   80.3$& $   64.2$& 0.27& 4& 2002-08-01 \\
$1455252-094118$& 11.80&  1.12& 0.10& 346.4& $  42.5$&  39.4& 300.7&  $  -1.3$& $   96.8$& $   67.9$& 0.29& 7& 2002-08-01 \\
$1513217-110217$&  8.60&  1.01& 0.10& 349.7& $  38.6$&   5.9& 305.2&  $   0.7$& $  -57.0$& $  -78.0$& 0.83& 7& 2002-07-29 \\
$1515571-074734$&  8.72&  1.14& 0.08& 353.2& $  40.6$&  10.6& 304.2&  $   1.1$& $  -14.5$& $  -23.9$& 0.37& 7& 2002-07-29 \\
$1516423-045646$&  8.61&  1.01& 0.10& 356.0& $  42.5$&   5.9& 303.0&  $   1.3$& $  -28.1$& $  -28.4$& 0.79& 7& 2002-07-29 \\
$1519212-100704$&  9.81&  1.04& 0.11& 351.9& $  38.3$&  11.7& 306.1&  $   0.8$& $  181.2$& $  167.0$& 0.67& 7& 2002-07-29 \\
$1522364-132713$&  9.66&  1.04& 0.13& 349.8& $  35.3$&  11.0& 308.4&  $   0.3$& $  101.9$& $   79.9$& 0.68& 7& 2002-07-29 \\
$1523209-172640$& 11.87&  1.11& 0.13& 346.8& $  32.2$&  39.9& 310.5&  $  -3.7$& $  150.6$& $  117.1$& 0.41& 7& 2002-08-01 \\
$1523355-130810$&  8.58&  1.14& 0.14& 350.3& $  35.4$&   9.7& 308.5&  $   0.4$& $   82.2$& $   61.8$& 0.62& 7& 2002-07-29 \\
$1525032-081538$&  9.57&  1.02& 0.09& 354.9& $  38.7$&   9.6& 306.4&  $   1.2$& $    6.4$& $    1.7$& 0.81& 7& 2002-07-29 \\
$1530594-115622$&  8.42&  1.05& 0.16& 352.9& $  35.0$&   6.4& 309.5&  $   0.8$& $  -36.7$& $  -48.6$& 0.78& 7& 2002-07-29 \\
$1531579+012959$&  8.63&  1.03& 0.05&   6.2& $  43.8$&   6.4& 303.2&  $   2.1$& $   27.3$& $   56.7$& 0.84& 7& 2002-07-29 \\
$1538463-023544$&  8.82&  1.05& 0.18&   3.3& $  39.9$&   7.7& 306.7&  $   2.0$& $   21.6$& $   43.2$& 0.72& 7& 2002-07-29 \\
$2037196-291738$&  7.92&  1.10& 0.06&  14.6& $ -34.7$&   6.2&  22.0&  $   1.0$& $ -133.7$& $  -82.2$& 0.90& 7& 2002-07-30 \\
$2037262-322359$& 12.12&  1.01& 0.07&  10.9& $ -35.4$&  30.5&  21.9&  $  -0.2$& $  214.6$& $  253.8$& 0.50& 7& 2002-08-01 \\
$2040025-255759$&  8.10&  1.08& 0.10&  18.7& $ -34.4$&   6.2&  22.8&  $   1.4$& $   70.5$& $  135.2$& 0.39& 6& 2002-07-31 \\
$2046335-283547$& 10.21&  1.05& 0.07&  16.1& $ -36.5$&  14.4&  24.1&  $   1.3$& $ -203.0$& $ -148.4$& 0.95& 7& 2002-07-15 \\
$2050020-345336$&  8.10&  1.02& 0.06&   8.5& $ -38.5$&   4.8&  24.4&  $   0.5$& $  -20.5$& $    9.0$& 0.79& 7& 2002-07-30/31 \\
$2053053-315513$&  8.39&  1.02& 0.11&  12.4& $ -38.6$&   5.6&  25.2&  $   0.7$& $   30.4$& $   71.9$& 0.68& 7& 2002-07-30 \\
$2059163-383923$& 10.64&  1.05& 0.04&   3.9& $ -40.7$&  18.1&  25.9&  $  -1.8$& $  142.4$& $  156.7$& 0.85& 7& 2002-07-18 \\
$2108046-335252$& 10.43&  1.03& 0.08&  10.5& $ -42.0$&  14.9&  28.2&  $  -0.2$& $  123.1$& $  156.5$& 1.01& 7& 2002-07-16 \\
$2112160-332351$& 11.59&  1.03& 0.11&  11.3& $ -42.8$&  25.5&  29.1&  $  -0.7$& $  -42.0$& $   -6.7$& 0.65& 7& 2002-08-01 \\
$2113139-295216$& 10.16&  1.02& 0.10&  16.1& $ -42.4$&  12.8&  29.7&  $   0.9$& $  -49.3$& $   -0.1$& 0.85& 7& 2002-07-18 \\
$2114412-301256$&  8.88&  1.07& 0.12&  15.7& $ -42.8$&   8.4&  30.0&  $   0.8$& $ -151.4$& $ -103.6$& 0.72& 7& 2002-07-30 \\
$2115336-353006$& 11.73&  1.03& 0.07&   8.5& $ -43.7$&  26.7&  29.5&  $  -1.8$& $   94.5$& $  121.1$& 0.60& 7& 2002-08-01 \\
$2117471-243233$& 11.92&  1.01& 0.05&  23.4& $ -42.2$&  27.4&  31.5&  $   3.2$& $   31.0$& $  100.7$& 0.56& 7& 2002-08-01 \\
$2125078-274709$& 10.65&  1.00& 0.10&  19.6& $ -44.6$&  14.5&  32.6&  $   1.3$& $   55.4$& $  112.1$& 0.85& 7& 2002-07-18 \\
$2130445-210034$&  9.01&  1.06& 0.04&  29.2& $ -44.1$&   8.6&  35.0&  $   2.1$& $   97.1$& $  179.4$& 0.64& 7& 2002-07-30 \\
$2131454-351351$& 11.65&  1.01& 0.07&   9.3& $ -47.0$&  24.0&  32.8&  $  -1.6$& $   85.5$& $  112.1$& 0.53& 7& 2002-08-01 \\
$2135183-203457$&  9.05&  1.10& 0.03&  30.3& $ -45.0$&  10.4&  36.1&  $   2.4$& $ -121.1$& $  -37.7$& 0.43& 5& 2002-07-30 \\
$2143212-363045$& 11.86&  1.01& 0.03&   7.5& $ -49.4$&  26.6&  34.9&  $  -2.7$& $  -42.0$& $  -21.7$& 0.69& 7& 2002-08-01 \\
$2147096-212258$&  8.90&  1.00& 0.04&  30.4& $ -47.8$&   6.6&  38.7&  $   1.7$& $   17.7$& $   96.6$& 0.72& 7& 2002-07-30 \\
$2149407-193625$& 11.69&  1.02& 0.03&  33.2& $ -47.8$&  26.1&  39.6&  $   4.8$& $  -75.5$& $    9.5$& 0.96& 7& 2002-08-01 \\
$2150144-211437$& 10.28&  1.06& 0.04&  30.9& $ -48.5$&  15.6&  39.4&  $   2.8$& $ -163.2$& $  -84.2$& 1.14& 7& 2002-07-16 \\
$2154471-224050$&  8.85&  1.04& 0.03&  29.3& $ -49.9$&   7.5&  40.2&  $   1.6$& $  -75.0$& $   -2.2$& 0.86& 7& 2002-07-30 \\
$2158199-340607$& 10.74&  1.07& 0.02&  11.4& $ -52.4$&  20.0&  38.4&  $  -1.2$& $   47.9$& $   75.8$& 0.61& 7& 2002-11-23 \\
$2208397-281212$& 10.92&  1.04& 0.02&  21.5& $ -54.1$&  19.1&  42.0&  $   0.6$& $   44.1$& $   93.4$& 0.83& 7& 2002-07-16 \\
$2208499-292337$&  9.66&  1.02& 0.02&  19.5& $ -54.3$&  10.1&  41.7&  $   0.6$& $  -19.5$& $   25.0$& 0.76& 7& 2002-07-30 \\
$2213521-163152$&  8.11&  1.01& 0.03&  40.8& $ -52.1$&   4.8&  46.0&  $   1.7$& $ -100.7$& $   -8.7$& 0.75& 7& 2002-09-26 \\
$2214268-230618$& 10.24&  1.10& 0.03&  30.5& $ -54.4$&  17.8&  44.5&  $   2.1$& $   -9.0$& $   58.4$& 0.67& 7& 2002-07-31 \\
$2218053-132605$&  8.45&  1.05& 0.04&  46.1& $ -51.7$&   6.4&  47.9&  $   2.3$& $   37.2$& $  139.3$& 0.79& 7& 2002-09-26 \\
$2222150-194151$&  8.13&  0.99& 0.03&  37.1& $ -55.1$&   4.4&  47.1&  $   1.4$& $    6.2$& $   84.7$& 0.78& 7& 2002-09-26 \\
$2226328-340408$& 11.56&  1.03& 0.01&  11.3& $ -58.2$&  24.7&  44.1&  $  -2.3$& $ -120.5$& $  -97.7$& 0.71& 7& 2002-08-01 \\
$2226495-391831$& 10.86&  1.01& 0.02&   1.5& $ -57.7$&  16.5&  42.7&  $  -2.7$& $   82.0$& $   84.3$& 0.68& 7& 2002-09-24 \\
$2228277-124852$&  8.76&  0.95& 0.05&  48.9& $ -53.7$&   5.1&  50.5&  $   2.0$& $   44.1$& $  145.7$& 0.81& 7& 2002-09-26 \\
$2230427-112154$&  8.08&  0.97& 0.06&  51.5& $ -53.4$&   3.9&  51.5&  $   1.8$& $  -38.6$& $   67.2$& 0.77& 7& 2002-09-26 \\
$2232532-085221$&  8.78&  1.00& 0.05&  55.5& $ -52.5$&   6.2&  52.8&  $   2.6$& $ -102.8$& $   11.0$& 0.81& 7& 2002-09-26 \\
$2234012-215834$&  8.61&  1.01& 0.03&  34.7& $ -58.4$&   5.9&  49.1&  $   1.3$& $  -45.7$& $   21.2$& 0.78& 7& 2002-07-30 \\
$2234288-272213$& 10.59&  0.98& 0.02&  24.5& $ -59.7$&  12.9&  47.7&  $   0.6$& $    9.2$& $   56.0$& 0.69& 7& 2002-09-24 \\
$2237398-262854$& 10.35&  1.04& 0.02&  26.4& $ -60.2$&  14.7&  48.6&  $   0.7$& $   19.3$& $   68.5$& 0.66& 7& 2002-07-31 \\
$2239225-250812$&  9.99&  1.10& 0.02&  29.2& $ -60.4$&  15.7&  49.4&  $   1.0$& $  -16.7$& $   37.0$& 0.46& 7& 2002-09-24 \\
$2240154-172843$&  8.61&  0.99& 0.04&  43.7& $ -58.3$&   5.5&  51.9&  $   1.6$& $ -116.4$& $  -34.6$& 0.79& 7& 2002-09-26 \\
$2243104+050503$&  9.04&  0.95& 0.07&  74.2& $ -45.3$&   5.7&  60.3&  $   3.6$& $  -55.5$& $   98.0$& 0.74& 7& 2002-09-26 \\
$2244223-324716$& 11.21&  1.03& 0.01&  13.5& $ -62.0$&  21.5&  48.1&  $  -1.9$& $   24.8$& $   48.2$& 0.66& 7& 2002-09-24 \\
$2246547-200802$&  9.27&  1.00& 0.03&  40.0& $ -60.7$&   7.9&  52.6&  $   1.5$& $  -89.7$& $  -19.5$& 0.89& 7& 2002-09-24 \\
$2249453-012622$&  8.55&  0.96& 0.10&  69.2& $ -51.1$&   4.7&  59.5&  $   2.7$& $  -42.0$& $   90.5$& 0.76& 7& 2002-09-26 \\
$2249497-273039$&  9.49&  1.12& 0.03&  25.0& $ -63.1$&  13.9&  50.9&  $   0.2$& $  -57.5$& $  -15.5$& 0.55& 7& 2002-07-30 \\
$2256121-204556$& 11.24&  0.99& 0.03&  40.3& $ -63.0$&  18.6&  54.4&  $   1.9$& $  -49.1$& $   15.8$& 0.66& 7& 2002-09-24 \\
$2257436+071647$&  8.63&  1.00& 0.07&  80.2& $ -45.9$&   5.8&  65.1&  $   3.8$& $  -61.5$& $   93.3$& 0.56& 7& 2002-09-26 \\
$2300455+042329$&  8.66&  1.00& 0.07&  78.4& $ -48.7$&   5.9&  64.6&  $   3.5$& $ -206.4$& $  -60.4$& 0.80& 7& 2002-09-26 \\
$2302578-352028$& 12.15&  0.99& 0.02&   6.6& $ -65.5$&  27.7&  50.9&  $  -4.4$& $   89.5$& $   98.0$& 0.52& 7& 2002-09-30 \\
$2303483+082823$&  8.30&  0.99& 0.05&  82.9& $ -45.9$&   4.8&  67.2&  $   3.3$& $  -23.0$& $  133.0$& 0.72& 7& 2002-09-26 \\
$2308467+081215$&  7.99&  0.95& 0.04&  84.2& $ -46.8$&   3.5&  68.4&  $   2.6$& $ -242.4$& $  -88.8$& 0.96& 7& 2002-09-26 \\
$2309567-325813$& 12.06&  0.95& 0.01&  11.8& $ -67.3$&  23.4&  53.1&  $  -2.9$& $  117.0$& $  132.4$& 0.36& 7& 2002-09-30 \\
$2310036+092345$&  8.18&  0.97& 0.05&  85.5& $ -46.0$&   4.2&  69.3&  $   3.0$& $  -55.8$& $  100.2$& 0.73& 7& 2002-09-26 \\
$2314166+051713$&  8.65&  0.99& 0.08&  83.4& $ -50.0$&   5.6&  68.5&  $   3.3$& $   42.8$& $  186.4$& 0.82& 7& 2002-09-26 \\
$2314268-261553$&  8.53&  1.03& 0.03&  29.7& $ -68.4$&   6.2&  56.5&  $   0.5$& $   23.4$& $   62.2$& 0.81& 7& 2002-07-30 \\
$2315227+052700$&  8.22&  1.02& 0.09&  83.9& $ -50.0$&   5.1&  68.8&  $   3.2$& $  -40.3$& $  103.1$& 0.59& 7& 2002-09-26 \\
$2318198+013515$&  8.11&  0.97& 0.04&  81.2& $ -53.6$&   4.0&  67.9&  $   2.4$& $ -156.6$& $  -25.2$& 0.81& 7& 2002-09-26 \\
$2319435-154611$& 11.24&  1.03& 0.03&  56.3& $ -65.9$&  21.8&  61.4&  $   3.1$& $  -82.0$& $   -7.3$& 0.57& 7& 2002-09-27 \\
$2321018+080136$&  9.25&  0.96& 0.10&  87.8& $ -48.6$&   6.5&  71.5&  $   3.9$& $  -36.0$& $  112.1$& 0.49& 7& 2002-09-26 \\
$2321265-242654$& 10.50&  1.14& 0.02&  35.4& $ -69.6$&  24.0&  58.6&  $  -0.1$& $   -0.1$& $   42.6$& 0.69& 7& 2002-07-15 \\
$2321309+082002$&  8.52&  1.00& 0.06&  88.2& $ -48.4$&   5.6&  71.7&  $   3.5$& $   -8.8$& $  140.2$& 0.63& 7& 2002-09-26 \\
$2321593+091110$&  9.03&  0.96& 0.04&  89.0& $ -47.7$&   5.9&  72.3&  $   3.7$& $  -36.2$& $  114.7$& 0.79& 7& 2002-09-26 \\
$2322353-314437$& 12.32&  1.01& 0.01&  14.2& $ -70.2$&  32.7&  56.1&  $  -4.3$& $  -93.9$& $  -78.2$& 0.57& 7& 2002-08-01 \\
$2324147-275034$& 11.46&  1.04& 0.02&  25.8& $ -70.7$&  24.9&  57.9&  $  -1.6$& $    1.5$& $   31.0$& 0.61& 7& 2002-09-28 \\
$2324366-133040$& 12.10&  1.06& 0.03&  62.5& $ -65.6$&  36.1&  63.4&  $   5.7$& $  -75.5$& $    4.7$& 0.64& 7& 2002-08-01 \\
$2324391+061649$&  8.65&  0.99& 0.10&  87.6& $ -50.5$&   5.7&  71.6&  $   3.4$& $ -125.1$& $   16.8$& 0.76& 7& 2002-09-26 \\
$2326237-250037$& 10.78&  1.00& 0.02&  34.5& $ -70.8$&  15.7&  59.4&  $   0.0$& $  -27.7$& $   11.2$& 0.81& 7& 2002-07-16/18 \\
$2329301-245810$& 10.46&  1.10& 0.02&  35.0& $ -71.5$&  20.1&  60.1&  $  -0.3$& $  -44.8$& $   -6.8$& 0.66& 7& 2002-07-15 \\
$2329548-205103$& 10.59&  1.00& 0.04&  47.2& $ -70.4$&  14.4&  61.8&  $   1.0$& $  -46.2$& $    6.3$& 0.71& 7& 2002-09-27 \\
$2330346-160747$& 11.30&  0.98& 0.03&  59.2& $ -68.3$&  17.9&  63.7&  $   2.3$& $  -65.5$& $    3.3$& 0.84& 7& 2002-07-19 \\
$2336339-240843$& 11.53&  1.03& 0.02&  38.8& $ -72.9$&  24.3&  61.9&  $  -0.5$& $  119.9$& $  158.1$& 0.72& 7& 2002-09-28 \\
$2337576+082014$&  8.16&  1.03& 0.12&  93.7& $ -50.3$&   5.2&  75.9&  $   3.2$& $  -36.0$& $  106.0$& 0.58& 7& 2002-09-26 \\
$2341204+002901$&  9.61&  1.03& 0.02&  88.7& $ -57.7$&  10.3&  73.0&  $   4.2$& $ -258.0$& $ -139.7$& 0.66& 7& 2002-09-27 \\
$2342433+130143$&  8.87&  1.00& 0.04&  98.2& $ -46.5$&   6.5&  79.4&  $   4.1$& $ -107.4$& $   44.4$& 0.85& 7& 2002-09-26 \\
$2343079-235827$& 11.35&  0.97& 0.02&  40.7& $ -74.3$&  17.8&  63.4&  $  -0.3$& $  -30.2$& $    5.8$& 0.82& 7& 2002-07-16 \\
$2345417-264456$& 10.38&  1.02& 0.02&  30.7& $ -75.4$&  14.0&  62.8&  $  -0.7$& $  -15.4$& $    9.6$& 0.78& 7& 2002-07-31 \\
$2346267-460845$&  9.52&  1.04& 0.01& 332.7& $ -67.0$&  10.3&  54.4&  $  -3.3$& $  -84.7$& $ -129.4$& 0.85& 7& 2002-09-28 \\
$2350361-200216$& 10.58&  1.07& 0.02&  56.7& $ -74.4$&  18.4&  66.5&  $   0.6$& $  -61.1$& $  -14.2$& 0.42& 7& 2002-09-27 \\
$2353194-205041$& 11.61&  0.97& 0.02&  55.2& $ -75.3$&  20.1&  66.8&  $   0.2$& $  -59.4$& $  -16.6$& 0.80& 7& 2002-07-15 \\
$2355226+013932$&  8.81&  1.00& 0.02&  95.6& $ -58.2$&   6.3&  76.9&  $   2.9$& $ -139.3$& $  -24.1$& 0.92& 7& 2002-09-26 \\
$2356129+053127$&  9.16&  1.09& 0.05&  98.6& $ -54.7$&  10.6&  78.9&  $   4.8$& $ -138.6$& $  -12.7$& 0.42& 7& 2002-09-27 \\
$2356135+121516$&  8.38&  0.98& 0.08& 102.4& $ -48.4$&   4.7&  82.3&  $   3.1$& $  -34.6$& $  109.1$& 0.76& 7& 2002-09-26 \\
$2356374-234712$& 11.46&  0.99& 0.02&  45.0& $ -77.2$&  20.8&  66.3&  $  -0.9$& $  -44.8$& $  -13.8$& 0.88& 7& 2002-07-18 \\
$2356542-190609$& 11.64&  1.03& 0.02&  62.7& $ -75.1$&  25.6&  68.3&  $   0.6$& $  -68.8$& $  -21.5$& 0.93& 7& 2002-08-01 \\
$2358327-202740$&  9.26&  1.01& 0.02&  59.0& $ -76.2$&   7.9&  68.1&  $   0.6$& $ -102.6$& $  -60.8$& 0.78& 7& 2002-09-27 \\
$2359305+094815$&  8.43&  0.97& 0.11& 102.3& $ -51.0$&   4.6&  81.9&  $   2.9$& $  -36.6$& $  107.1$& 0.71& 7& 2002-09-26 \\

\enddata
\end{deluxetable}
\end{document}